\documentclass[twocolumn,prd,superscriptaddress,showpacs,
floatfix,preprintnumbers,nofootinbib]{revtex4}

\usepackage{epsfig}
\usepackage{amsmath,bm}
\usepackage{color}

\newcommand{\I}{{\rm i}}

\newcommand{\sF}{{\sf F}}

\newcommand{\sH}{{\sf H}}
\newcommand{\sM}{{\sf M}}
\newcommand{\sN}{{\sf N}}

\def\beq{\begin{equation}}
\def\eeq{\end{equation}}

\definecolor{purple}{rgb}{0.63,0.13,0.94}
\definecolor{red}{rgb}{1.0,0.0,0.0}
\definecolor{green}{rgb}{0.0,1.0,0.0}
\definecolor{blue}{rgb}{0.0,0.0,1.0}
\definecolor{orange}{rgb}{0.8,0.6,0,0}
\definecolor{magenta}{rgb}{0.8,0.0,0.6}
\definecolor{white}{rgb}{1.0,1.0,1.0}
\definecolor{black}{rgb}{0.0,0.0,0.0}

\begin{document}

\title{Linearized flavor-stability analysis of dense neutrino streams}

\author{Arka Banerjee}
\affiliation{Tata Institute of Fundamental Research, Homi Bhabha
Road, Mumbai 400005, India}

\author{Amol Dighe}
\affiliation{Tata Institute of Fundamental Research, Homi Bhabha
Road, Mumbai 400005, India}

\author{Georg Raffelt}
\affiliation{Max-Planck-Institut f\"ur Physik
(Werner-Heisenberg-Institut), F\"ohringer Ring 6, 80805 M\"unchen,
Germany}

\date{\today}


\begin{abstract}
Neutrino-neutrino interactions in dense neutrino streams, like those
emitted by a core-collapse supernova, can lead to self-induced neutrino
flavor conversions. While this is a nonlinear phenomenon,
the onset of these conversions can be examined through a
standard stability analysis of the linearized equations of motion.
The problem is reduced to a linear eigenvalue equation that
involves the neutrino density, energy spectrum, angular
distribution, and matter density. In the single-angle case, we
reproduce previous results and use them to identify two generic
instabilities: The system is stable above a cutoff density
(``cutoff mode''), or can approach an asymptotic instability
for increasing density (``saturation mode''). We
analyze multi-angle effects on these generic types of instabilities
and find that even the saturation mode is suppressed at large
densities. For both types of modes, a given multi-angle spectrum
typically is unstable when the neutrino and electron densities are
comparable, but stable when the neutrino density is much smaller or
much larger than the electron density. The role of an instability in
the SN context depends on the available growth time and on the range
of affected modes. At large matter density, most modes are
off-resonance even when the system is unstable.
\end{abstract}

\preprint{MPP-2011-81, TIFR/TH/11-30}


\maketitle

\section{Introduction}
\label{sec:intro}

Neutrino flavor oscillations in a supernova (SN) are strongly
suppressed by matter effects~\cite{Wolfenstein:1977ue} until the
neutrinos pass through the usual MSW region~\cite{Mikheev:1986gs,
Mikheev:1986if, Dighe:1999bi, Kuo:1989qe} far out in the envelope of
the collapsing star. However, neutrino-neutrino
interactions~\cite{Pantaleone:1992eq, Sigl:1992fn}, through a flavor
off-diagonal refractive index, can trigger self-induced flavor 
conversions~\cite{Samuel:1993uw, Kostelecky:1993dm,
Kostelecky:1995dt, Samuel:1996ri, Sawyer:2005jk, Sawyer:2008zs}.
This collective effect tends to occur between the neutrino sphere
and the MSW region and can lead to strongly modified neutrino
spectra, showing features such as spectral swaps and
splits~\cite{Duan:2006an, Raffelt:2007cb, Duan:2007fw, Fogli:2007bk,
Fogli:2008pt, Dasgupta:2009mg}; for a review see
Ref.~\cite{Duan:2010bg}. The overall scenario, supported by
heuristic arguments and numerical examples, is that deep inside the
SN core, the system performs ``synchronized oscillations'' with an
extremely small amplitude, i.e.\ every neutrino remains essentially
stuck in its initial flavor eigenstate. As the neutrinos stream
outwards, there is a sharp onset radius where ``bimodal''
oscillations begin: Some ranges of modes start pendulum-like
oscillations~\cite{Samuel:1996ri, Hannestad:2006nj, Duan:2007mv,
Raffelt:2011yb}, exchanging their flavor content with each other
without affecting the flavor content of the overall system.

This scenario engenders a crucial simplification for the treatment
of neutrino transport in SN simulations. At high densities, where
neutrinos collide frequently, it is enough to solve the transport
equations for each flavor separately, ignoring oscillations
entirely. On the other hand, flavor conversions at larger distances
can be treated ignoring neutrino collisions and absorption, i.e.\ as
a pure propagation problem. So collisions and flavor oscillations
are phenomena that are assumed to be taking place in different
regions of the star and can be treated independently. When the
radial distance where bimodal oscillations begin is far away from
the SN core, this assumption is valid and the flavor conversions do
not affect the SN dynamics. Recent studies dedicated to the SN
accretion phase, under simplifying assumptions, once more confirm
this picture~\cite{Chakraborty:2011gd, Dasgupta:2011jf}.

However, what is missing is a systematic approach to decide, without
solving the equations of motion, if self-induced flavor conversions
occur for given neutrino spectra (flavor-dependent energy and
angular distribution), overall neutrino density, and matter density.
Formal stability criteria exist only in the ``single-angle
approximation'' where it is assumed that all neutrinos feel the same
neutrino-neutrino refractive effect. In this case the analytic
pendulum solution has been found and its existence and parameters
can be calculated from the neutrino spectrum and density
alone~\cite{Dasgupta:2009mg}.

On the other hand, the current-current nature of the low-energy
weak-interaction Hamiltonian implies that neutrinos in the
background of an anisotropic neutrino flux experience a refractive
effect that strongly depends on direction. For some energy spectra,
these ``multi-angle effects'' have little impact, whereas in other
cases they completely change the solution. A SN neutrino flux with a
vanishing net $\nu_e$ flux is unstable everywhere and shows quick
multi-angle decoherence~\cite{Raffelt:2007yz}. If the $\nu_e$ flux
is large enough, this effect is avoided if the energy spectrum is
simple~\cite{EstebanPretel:2007ec}. More complicated spectra can be
unstable even for a large $\nu_e$ flux at any density in the
single-angle case, but the instability is suppressed by multi-angle
effects~\cite{Raffelt:2008hr, Duan:2010bf}. In addition, the
presence of ordinary matter causes a multi-angle suppression of the
bimodal instability~\cite{EstebanPretel:2008ni}. On the other hand,
it was claimed that for nontrivial angular distributions, as may
exist when different flavors are emitted from very different
neutrino spheres, there is a novel multi-angle
instability~\cite{Sawyer:2005jk}. Deviations from the usually
assumed cylindrical symmetry may also have an important
influence~\cite{Sawyer:2008zs}.

Although our problem is nonlinear and therefore would seem
intractable, noting that an instability must occur in order for the
onset to take place leads to a surprising simplification. In the
dense SN matter well inside of the MSW region, the matter effect is
so large that neutrino propagation eigenstates are essentially
identical with flavor eigenstates. This means that in the
weak-interaction basis, the flavor matrices of occupation numbers
are almost perfectly diagonal, allowing us to linearize the
equations of motion (EoMs) in terms of the small off-diagonal
elements. An instability is equivalent to some of these small
elements starting to grow exponentially. All we need to do is
linearize the EoMs, perform a Fourier analysis, and seek
exponentially growing solutions of the relevant eigenvalue equation.
This is simple even in the multi-angle situation. Almost certainly
this approach can be extended to cases without the usually assumed
cylindrical symmetry around the radial direction.

Studying the stability of a strongly coupled and nonlinear system in
the small-amplitude limit is a standard technique. In the present
context it was put forth in Ref.~\cite{Sawyer:2008zs}. However, the
method was carried only to schematic cases of a small number
of neutrino momentum modes, leaving open how to apply it to
realistic situations.

While a stability analysis provides crucial insight, we stress that
it alone is not enough to assess the impact in the SN context. Only
a small range of modes may participate in the bimodal oscillation or
the growth rate may be too small on the available time scale. (Of
course, both the growth rate and the location of the resonance on
the energy-angle spectrum are found by solving the eigenvalue
equation.) For example, the classic single-angle case with an
initial Fermi-Dirac spectrum of only $\nu_e$ and $\bar\nu_e$ is
always unstable and the usual concept of a synchronization radius
does not apply. However, for large neutrino densities, the growth
rate is small and only a narrow range of infrared modes is affected.
A ``visible'' effect arises only at a quasi-onset radius where the
growth rate begins to compete with the overall evolution time scale
and the resonance begins to move into the main part of the spectrum.
We will here largely avoid such issues and concentrate on setting up
the method and discussing simple but informative schematic cases.
Applying this method in a realistic SN context will be left to
future work.

Our study in Sec.~\ref{sec:eoms} leads to the linearized equations
of motion at a large distance from the neutrino source in the
two-flavor case, with an azimuthally symmetric neutrino emission. In
Sec.~\ref{sec:single}, we present the stability analysis in the
single-angle approximation, and illustrate it in
Sec.~\ref{sec:single-ex} with the examples of box spectra where the
results may be understood analytically. In Sec.~\ref{sec:nosleep} we
point out some special features of
realistic spectra that do not vanish at low energies.
Sections~\ref{sec:multi} and \ref{sec:multi-ex} demonstrate how the
single-angle results are modified by the inclusion of multi-angle
and matter effects. The latter one also analyzes a realistic
SN spectrum using the insights obtained from the box spectra. In
Sec.~\ref{sec:novel}, we show that a multi-angle spectrum
with a zero crossing is unstable in both hierarchies
if the lepton asymmetry is small.
In Sec.~\ref{sec:conclusions} we conclude with a brief summary of our
findings and an outlook on future directions.

\section{Equations of motion}
\label{sec:eoms}

\subsection{Effective Hamiltonian}
\label{hamiltonian}

We write the EoMs in terms of $n_{\rm F}{\times}\,n_{\rm F}$
matrices of occupation numbers~\cite{Sigl:1992fn,Dolgov:1980cq},
where $n_{\rm F}$ is the number of flavors. We denote these matrices
by $\varrho_{E,{\bf v}}$, where the velocity vector ${\bf v}$ with
$|{\bf v}|=1$ describes the direction of motion and the energy $E$
is taken to be positive for neutrinos and negative for
antineutrinos. Also, while the diagonal entries are equal to the
occupation numbers for neutrinos, they are the negative occupation
numbers for antineutrinos. In the context of a Boltzmann collision
equation for mixed neutrinos, one uses positive occupation numbers
in both cases and describe each mode by its momentum ${\bf p}$
\cite{Sigl:1992fn}. Our choice of signs, however, allows us to
include neutrinos and antineutrinos on the same footing and we will
never have to distinguish between them: The antineutrino spectrum is
simply a continuation of the neutrino spectrum to negative energies.
In the language of flavor polarization vectors, our convention
agrees with the neutrino flavor isospin
construction~\cite{Duan:2006an}.

The EoMs for the time evolution in a homogeneous medium are
\begin{equation}
\I\partial_t\varrho_{E,{\bf v}}=[\sH_{E,{\bf v}},\varrho_{E,{\bf v}}]\,.
\end{equation}
We use sans-serif letters for matrices in flavor space. The
Hamiltonian matrix is
\begin{eqnarray}
\sH_{E,{\bf v}}&=&\frac{\sM^2}{2E}+\sqrt{2}\,G_{\rm F}\sN_\ell
\\
&+&\sqrt{2}\,G_{\rm F}
\int_{-\infty}^{+\infty}dE'\int d{\bf v}' \frac{E'^2}{(2\pi)^3}
\,\varrho_{E',{\bf v}'}(1-{\bf v}\cdot{\bf v}')
\nonumber
\end{eqnarray}
where $\sM^2$ is the neutrino mass-squared matrix and $\sN_\ell$ the
matrix of net charged-lepton densities which in the flavor basis is
$\sN_\ell={\rm diag}(n_{e}{-}n_{\bar e},n_{\mu}{-}n_{\bar\mu},
n_{\tau}{-}n_{\bar\tau})$. In an isotropic medium, the ${\bf
v}\cdot{\bf v}'$ term drops out and the neutrino-neutrino term has
the same structure as the matter term: The phase-space integral over
$\varrho_{E,{\bf v}}$ amounts to the difference between neutrino and
antineutrino densities. In the presence of macroscopic matter
fluxes, there would also be current contributions in the matter
term~\cite{EstebanPretel:2008ni}.

\subsection{Azimuthal symmetry}
\label{azimuthal}

Henceforth we assume azimuthal symmetry around some preferred
direction, usually the radial direction in the SN case. The
azimuthal integration provides
\begin{equation}
1-{\bf v}\cdot{\bf v}'\to 1-{v}\cdot{v}' \; ,
\end{equation}
where $v$ and $v'$ are the components of ${\bf v}$ and ${\bf v}'$,
respectively, along the symmetry direction. Thus $v=\cos\vartheta$
with $\vartheta$ the trajectory angle relative to the symmetry
direction.

Following the Appendix of Ref.~\cite{EstebanPretel:2007ec} we
consider the EoMs expressed in terms of the radial coordinate. We
introduce an arbitrary sphere with radius $R$ that we call neutrino
sphere where we specify the inner boundary condition for neutrinos
that are assumed to stream only outward. Every angular mode is
described by its emission angle $\vartheta_R$ relative to the radial
direction at that sphere (Fig.~3 of Ref.~\cite{Duan:2010bg}) in
terms of the variable $u=\sin^2\vartheta_{R}$ which lies in the
range $0\leq u\leq 1$. The $u$ variable has the property that the
modes are uniformly distributed on $0\leq u\leq 1$ if the emission
at the neutrino sphere is isotropic into space in analogy to
blackbody emission.

At radius $r$, the radial velocity of a mode with angular label $u$
is
\begin{equation}
v_{u,r}=\sqrt{1-\frac{R^2}{r^2}\,u}\,.
\end{equation}
In analogy to Ref.~\cite{EstebanPretel:2007ec} we introduce the
matrices
\begin{equation}
{\sf\Phi}_{E,u,r}=\frac{r^2E^2}{2\pi}\,\varrho_{E,u,r}\,,
\end{equation}
where we have included a factor $4\pi r^2$, so that the integrated
quantity
\begin{equation}
{\sf\Phi}_{r}=\int_{-\infty}^{+\infty}dE\int_0^1du\,{\sf\Phi}_{E,u,r}
\end{equation}
represents the flux through a sphere of radius $r$ whose trace is
conserved.

The EoMs for the flux matrices as a function of radial coordinate
are
\begin{equation}
\I\partial_r{\sf\Phi}_{E,u,r}=[\sH_{E,u,r},{\sf\Phi}_{E,u,r}]
\end{equation}
with the Hamiltonian
\begin{eqnarray}
\sH_{E,u,r}&=&\left(\frac{\sM^2}{2E}+
\sqrt{2}\,G_{\rm F}\sN_\ell\right)\,\frac{1}{v_{u,r}}
\nonumber\\
&+&\frac{\sqrt{2}\,G_{\rm F}}{4\pi r^2}
\int_0^1 du'
\left(\frac{1}{v_{u,r}v_{u',r}}-1\right)\,{\sf\Phi}_{u',r}\,,
\end{eqnarray}
where ${\sf\Phi}_{u,r}=
\int_{-\infty}^{+\infty}dE\,{\sf\Phi}_{E,u,r}$.

\subsection{At a large distance from source}
\label{large-dist}

We are interested in the evolution far away from the neutrino sphere
where the flavor conversions are expected to begin. Therefore, we
use the expansion
\begin{equation}
v_{u,r}^{-1}=1+\frac{u}{2}\,\frac{R^2}{r^2}\,.
\end{equation}
Moreover, we introduce the dimensionless matrices ${\sf L}={\sf
N}_\ell/(n_{e}-n_{\bar e})$ and ${\sf
F}_{E,u,r}={\sf\Phi}_{E,u,r}/\Phi_{\bar\nu_e}(R)$. Note that we
normalize the charged-lepton density to the local net electron
density, whereas the neutrino flux matrices are normalized to the
total $\bar\nu_e$ flux $\Phi_{\bar\nu_e}(R)$ at the neutrino sphere. 
If we use the flavor
basis, with these normalizations we have ${\sf L}^{ee}=1$ and
$\int_{-\infty}^0 dE\,\int_0^1du\,{\sf F}^{ee}_{E,u,r}=-1$ for all
$r$ where oscillations have not yet begun.

We also introduce the coefficients with dimension of inverse energy
\begin{eqnarray}
\tilde\lambda_r&=&\sqrt{2}\,G_{\rm F}\left[n_{e}(r)-n_{\bar e}(r)\right] \; ,
\nonumber\\
\mu_R&=&\frac{\sqrt{2}\,G_{\rm F}\Phi_{\bar\nu_e}(R)}{4\pi R^2}\,.
\end{eqnarray}
In terms of these coefficients, we have
\begin{equation}
\I\partial_r{\sf F}_{E,u,r}=[\sH_{E,u,r},{\sf F}_{u,r}] \; ,
\end{equation}
with
\begin{eqnarray}
\sH_{E,u,r}&=&\left(\frac{\sM^2}{2E}+\tilde\lambda_r{\sf L}\right)
\,\left(1+\frac{u}{2}\,\frac{R^2}{r^2}\right)
\nonumber\\
&+&\mu_R\,\frac{R^4}{r^4} \int_0^1 du'\,\frac{u+u'}{2}\,\sF_{u',r}
\label{H-lowest-order}
\end{eqnarray}
as the Hamiltonian at the lowest-order in $(R/r)$, with $\sF_{u,r} =
\int_{-\infty}^{+\infty} dE \, \sF_{E,u,r}$. The first line on the
right hand side of Eq.~(\ref{H-lowest-order}) is the ``vacuum plus
matter'' Hamiltonian $\sH_{E,u,r}^{\rm vac+mat}$ while the second
line is the neutrino-neutrino Hamiltonian $\sH_{E,u,r}^{\nu\nu}$.

\subsection{Two-flavor case}
\label{two-flavor}

For the rest of this paper, we restrict ourselves to the two-flavor
scenario, with flavors $e$ and $x$, and we introduce the variable
$\omega=|\Delta m^2|/2E$, the vacuum oscillation frequency, to
describe the different modes. In the context of flavor oscillation
physics, $\omega$ is a much more natural variable to describe the
neutrino spectrum than the energy $E$. Note that since $E$ is taken
to be negative for antineutrinos, they are represented by negative
$\omega$ values. Since the trace of the Hamiltonian does not
contribute to the time evolution, we write
\begin{eqnarray}
\frac{{\sf M}^2}{2E}&=& \pm \frac{\omega}{2}\,
\begin{pmatrix}
\cos{2\theta}&\sin{2\theta}\\
-\sin{2\theta}&-\cos{2\theta}
\end{pmatrix}\,,
\nonumber\\
\tilde\lambda_r{\sf L}&=&\frac{\tilde\lambda_r}{2}\,
\begin{pmatrix}1&0\\0&-1\end{pmatrix}\,,
\end{eqnarray}
in the flavor basis, after removing a term proportional to the unit
matrix. We take the mixing angle $\theta$ to lie in the
first octant $0<\theta<\pi/4$. In this case the $+(-)$ sign stands
for inverted (normal) mass hierarchy. In the following discussion,
we shall consider inverted hierarchy. For obtaining results with
normal hierarchy, we will have to multiply the $\omega$ term by a
factor of $-1$.

The flux matrices at the neutrino sphere are
\begin{equation}
{\sf F}_{\omega,u,R}=
\begin{pmatrix}\phi^e_{\omega,u}&0\\0&\phi^x_{\omega,u}\end{pmatrix}\,,
\end{equation}
where the $\phi_{\omega,u}$ are differential fluxes in the variables
$\omega$ and $u$.  The normalization of ${\sf F}$ used here implies
that $\int_{-\infty}^0d\omega\,\int_0^1du\,\phi^e_{\omega,u}=-1$.
Note that $\phi_{\omega,u}$ for antineutrinos ($\omega <0$)
corresponds to the negative of their occupation numbers. Finally, in
the flavor basis we write
\begin{equation}
{\sf F}_{\omega,u,r}=\frac{{\rm Tr}\,{\sf F}_{\omega,u,r}}{2}
+\frac{g_{\omega,u}}{2}\,{\sf S}_{\omega,u,r} \; ,
\end{equation}
where $g_{\omega,u}=\phi^e_{\omega,u}-\phi^x_{\omega,u}$ is the
usual difference spectrum, except that it is now also differential
with regard to the direction variable $u$. The initial conditions at
the neutrinosphere for the Hermitian matrix ${\sf S}_{\omega,u,r}$
are
\begin{equation}
{\sf S}_{\omega,u,R} =
\begin{pmatrix} 1&0\\0&-1\end{pmatrix}\,.
\end{equation}
It satisfies the EoMs
\begin{equation}
\I\partial_r{\sf S}_{\omega,u,r}=[\sH_{E,u,r},{\sf S}_{\omega,u,r}]
\end{equation}
with the neutrino-neutrino part of the Hamiltonian
\begin{equation}
\sH_{\omega,u,r}^{\nu\nu}=
\mu_r \int_0^1 du'\,(u+u')
\int_{-\infty}^{+\infty}d\omega'\,\frac{g_{\omega'u'}}{2}\,
{\sf S}_{\omega',u',r}
\end{equation}
and
\begin{equation}
\mu_r=\mu_R\,\frac{R^4}{2r^4}\,.
\end{equation}
The effective neutrino-neutrino interaction energy declines with
$r^{-4}$.

\subsection{In a co-rotating frame}
\label{co-rotating}

We go to a rotating frame where the common matter term drops out and
where the vacuum term oscillates quickly, averaging the off-diagonal
term to zero~\cite{EstebanPretel:2008ni}. Moreover, in the large-$r$
limit we ignore a small radius-dependent shift of $\omega$. Then we
find
\begin{equation}
{\sf H}_{\omega,u,r}^{\rm vac+mat}=
\frac{\omega+u\lambda_r}{2}
\begin{pmatrix}1&0\\0&-1\end{pmatrix}\,,
\end{equation}
where
\begin{equation}
\lambda_r=\tilde\lambda_r\,\frac{R^2}{2r^2}
=\sqrt{2}\,G_{\rm F}\left[n_{e}(r)-n_{\bar e}(r)\right]\,\frac{R^2}{2r^2}
\label{lambdar-def}
\end{equation}
encodes the net matter effect. Note that the $1/r^2$ variation here
has nothing to do with the matter density profile, the effect of the
latter appears through the $n_e(r)$ term.

Next we write the ${\sf S}$ matrices in components in the flavor
basis
\begin{equation}
{\sf S}_{\omega,u,r}=
\begin{pmatrix}s_{\omega,u,r}&S_{\omega,u,r}\\
S_{\omega,u,r}^*&-s_{\omega,u,r}\end{pmatrix}\,,
\end{equation}
where $s_{\omega,u,r}$ is the $r$-dependent swap factor. It
specifies how much the flavor content of the given mode has been
swapped relative to the initial condition. We have the normalization
$s_{\omega,u,r}^2+|S_{\omega,u,r}|^2=1$. Likewise,
\begin{equation}
{\sf H}_{\omega,u,r}=
\begin{pmatrix}h_{\omega,u,r}&H_{\omega,u,r}\\
H_{\omega,u,r}^*&-h_{\omega,u,r}\end{pmatrix}\,.
\end{equation}
Then the EoM for the off-diagonal component is
\begin{equation}
\I\partial_r S_{\omega,u,r}=2\left(h_{\omega,u,r}S_{\omega,u,r}
-s_{\omega,u,r}H_{\omega,u,r}\right)\,.
\end{equation}
The components of the Hamiltonian matrix are explicitly
\begin{eqnarray}
h_{\omega,u,r}&=&\frac{\omega+u\,\lambda_r}{2}
\nonumber\\
&+&\frac{\mu_r}{2} \int_0^1 du'\,(u+u')
\int_{-\infty}^{+\infty}d\omega'\,g_{\omega'u'}\,
s_{\omega',u',r}\,,
\nonumber\\* H_{\omega,u,r}&=& \frac{\mu_r}{2}
\int_0^1 du'\,(u+u')
\int_{-\infty}^{+\infty}d\omega'\,g_{\omega'u'}\,
S_{\omega',u',r}\,.
\nonumber\\
\end{eqnarray}

In the absence of all interactions, the rotation-averaged EoM is
\begin{equation}
\I\partial_r S_{\omega,u,r}=\omega\,S_{\omega,u,r}\,,
\end{equation}
implying the free precession solution
\begin{equation}
S_{\omega,u,r}=e^{-\I\omega(r-R)}\,S_{\omega,u,R}\,.
\end{equation}

\subsection{Small-amplitude expansion}
\label{expansion}

Henceforth we drop the explicit subscript $r$ to denote the
$r$-dependence of all quantities. Moreover, we drop the limits of
integration which are always as above. In the small-amplitude case
we have $s_{\omega,u}=1$. This simplifies in particular the diagonal
Hamiltonian term which is
\begin{equation}
h_{\omega,u}=\frac{\omega+u\,\lambda}{2}
+\frac{\mu}{2} \int du'\,(u+u')
\int d\omega'\,g_{\omega',u'}\,.
\end{equation}
In the neutrino-neutrino term, the integral which involves $\int du'
u'\ldots$ is a constant that does not depend on $\omega$ or $u$ and
therefore amounts to a shift of all frequencies, i.e.\ yet another
rotating frame. Once more we can drop this term and are left with
\begin{equation}
h_{\omega,u}=\frac{\omega+u\,(\lambda+\epsilon\mu)}{2}\,.
\end{equation}
Here,
\begin{equation}
\epsilon=\int du\,d\omega\,g_{\omega,u}
\end{equation}
quantifies the ``asymmetry'' or ``total lepton number'' of the
neutrino spectrum, normalized to the total $\bar\nu_e$ flux. The
EoMs are then explicitly
\begin{eqnarray}\label{eq:smallEoM}
\I\partial_r S_{\omega,u}&=&\left[\omega+u(\lambda+\epsilon\mu)\right]S_{\omega,u}
\nonumber\\
&-&\mu \int du'\,d\omega'\,(u+u')\,g_{\omega'u'}\,S_{\omega',u'}\,.
\label{stability-eom}
\end{eqnarray}
This is the linearized form of the EoMs and provides the starting
point for the stability analysis.

At this point it may be useful to recapitulate the elements that
have gone into this analysis. Besides the small-amplitude
approximation $|S_{\omega,u}|\ll 1$, we have taken the neutrinos to
be far away from the neutrino sphere, $R/r\ll 1$. At the same time,
they have not yet reached the MSW resonance region, so that the
ordinary matter effect is large and the effective mixing angle in
matter is small. We have also assumed that the vacuum mixing angle
is so small that we may approximate $\cos\theta=1$, but it is
trivial to accommodate another choice. We have assumed that the
fast-rotating off-diagonal component of the Hamiltonian matrix
caused by the mismatch between the mass and flavor directions
averages to zero, so the only off-diagonal contribution of the
Hamiltonian is provided by the neutrinos themselves. In numerical
simulations, this fast-rotating component provides the initial
disturbance to kick-start exponentially growing modes. Here,
however, we do not ask how the instability gets started, we only ask
for the existence of exponentially growing modes.

\subsection{Eigenvalue equation}
\label{sec:fourier}

The stability analysis determines if the small quantities
$S_{\omega,u}$ grow exponentially with $r$. This is achieved by
writing $S_{\omega,u}$ as
\begin{equation}
S_{\omega,u}=Q_{\omega,u}\,e^{-\I\Omega r} \; ,
\label{fourier}
\end{equation}
where both $Q_{\omega,u}$ and $\Omega$ are in general complex
numbers. A purely real solution for $\Omega$ would imply that all
modes precess with a common frequency. A complex solution $\Omega
\equiv \gamma + i \kappa$, with $\kappa
>0$, would indicate an exponentially increasing $S_{\omega,u}$, i.e.,
an instability. On the other hand, $\kappa<0$ would
indicate that the solution decreases exponentially toward the
asymptotic solution $S_{\omega,u}=0$.

In terms of $Q_{\omega,u}$, the EoM becomes
\begin{eqnarray}
(\omega + u \bar\lambda - \Omega) Q_{\omega,u}&=&
\mu \int du'\,d\omega'\,(u+u')\,g_{\omega'u'}\,Q_{\omega',u'}\,, \nonumber \\
\label{fourier-eom}
\end{eqnarray}
where $\bar\lambda \equiv \lambda + \epsilon \mu$. This may
be looked upon as an eigenvalue equation for $Q_{\omega,u}$, which
is a vector in the function space on the $\omega$-$u$ plane, with
the eigenvalue $\Omega$.
The eigenfunction $Q_{\omega,u}$ implicitly
carries a label $\Omega$ because usually for every $\Omega$ there
exists a different $Q_{\omega,u}$.
Note that if $Q_{\omega,u}$ and
its corresponding $\Omega$ satisfies Eq.~(\ref{fourier-eom}), so
does $Q^*_{\omega,u}$ with the eigenvalue $\Omega^*$. This implies
that for each complex solution for $\Omega = \gamma + i \kappa$,
there exists another solution $\Omega = \gamma - i\kappa$. Thus, the
exponentially increasing and decreasing solutions always appear in
pairs.

\section{Single-angle stability analysis}
\label{sec:single}

\subsection{The consistency conditions}
\label{sec:consistency}

Many important features of collective neutrino oscillation
phenomena can be understood in a simplified model where only a
single angular mode $u=u_0$ is occupied, i.e., all neutrinos are
assumed to be emitted from the neutrino sphere at a fixed angle
relative to the radial direction. We therefore first perform the
stability analysis with this assumption and later extend it to the
multi-angle case. In the simplest schematic SN model, the neutrino
sphere at radius $R$ is pictured as a blackbody source without limb
darkening, in which case the angular distribution is such that $u$
is uniformly distributed on the interval $0\leq u\leq 1$. It is
natural to represent this case in the single-angle approximation by
$u_0=1/2$. For the time being, however, we keep $u_0$ as a
free parameter.

Since $u=u_0$, the term $u\bar\lambda$ in the EoM corresponds to a
common precession for all modes. We therefore can go to a basis
rotating with frequency $u_0 \bar\lambda$, in which
Eq.~(\ref{stability-eom}) becomes
\begin{equation}
\I\partial_r S_{\omega}=\omega\,S_{\omega}
-2 u_0 \mu
\int d\omega'\,g_{\omega'}\,
{S}_{\omega'}\,.
\label{eom-single}
\end{equation}
The single-angle approximation is then equivalent to saying that
all the neutrinos feel the same refractive effect due to the other neutrinos.
Requiring the solution to be of the form
$S_{\omega}=Q_\omega\,e^{-\I\Omega r}$ gives
\begin{equation}
(\omega-\Omega)\,Q_{\omega}=
2 u_0 \mu \int d\omega'\,g_{\omega'}Q_{\omega'}\,.
\label{consistency-single}
\end{equation}
This is the single-angle form of the eigenvalue equation in
Eq.~(\ref{fourier-eom}).

For the l.h.s.\ of the eigenvalue equation to be independent of
$\omega$ like the r.h.s., we must have
\begin{equation}
Q_\omega \propto\frac{1}{\omega-\Omega} \;
\end{equation}
and therefore
\begin{equation}\label{eq:singleangleeigenvalue}
\mu^{-1}= 2 u_0 \int  d\omega\,\frac{g_{\omega}}{\omega-\Omega}\; .
\end{equation}
For an instability, this equation should have a complex root
$\Omega= \gamma + \I \kappa$. Then, splitting the equation into real
and imaginary part, one obtains the two equations
\begin{eqnarray}
(2 u_0 \mu)^{-1}&=& \int d\omega\,
g_{\omega}\frac{\omega-\gamma}{(\omega-\gamma)^2+\kappa^2}\,,
\label{consistency-1} \\
0&=& \int  d\omega\,
g_{\omega}\frac{\kappa}{(\omega-\gamma)^2+\kappa^2}\,.
\label{consistency-2}
\end{eqnarray}
Eqs.~(\ref{consistency-1}) and (\ref{consistency-2}) are the
conditions that must be simultaneously satisfied by $\gamma$ and the
positive quantity $\kappa^2$. If a solution exists, we automatically
have a pair of solutions $\Omega=\gamma+\pm\I|\kappa|$, of which one
corresponds to an instability that will grow at the rate
$e^{|\kappa| t}$.

When an instability occurs, $|Q_\omega|^2$ takes the form of 
a Lorentzian centered at $\omega = \gamma$ and half-width characterized 
by $\kappa$. Thus, the solutions for $\gamma$ and $\kappa$ tell us the 
range of $\omega$-modes which are significantly affected.
We therefore present our results in the form of the parameters 
$\gamma$ and $\kappa$.
Note that significant flavor transformations take place only if
$g_\omega$ is significant in this range.

In the form of Eqs.~(\ref{consistency-1}) and
(\ref{consistency-2}), these results were previously derived in
Ref.~\cite{Dasgupta:2009mg} where the full solution was provided,
not only the small-amplitude expansion. The exponentially growing
and shrinking solutions correspond to the ``flavor pendulum'' near
its inverted position close to the beginning or end of a full swing.
A purely real $\Omega$ (no exponential growth) corresponds to the
pure precession
mode~\cite{Raffelt:2007cb,Duan:2007fw,Raffelt:2011yb}. While
Eqs.~(\ref{consistency-1}) and (\ref{consistency-2}) are not new,
the linearized analysis illustrates the origin of the Lorentz
denominator in the eigenfunction~$Q_\omega$.

A pure precession mode is described by a real eigenvalue
$\Omega=\gamma$. In this case $Q_\omega$ becomes singular for
$\omega=\gamma$ and the small-amplitude expansion is not
self-consistent if $g(\gamma)\neq 0$. The full non-singular
self-consistency relation for the pure precession mode, without the
small-amplitude approximation, was provided in
Ref.~\cite{Raffelt:2007cb} and of course agrees in the appropriate
limit. The linearized equations are useful to study the presence of
instabilities, but not necessarily to study the pure precession
solutions.

\subsection{A single spectral zero crossing}
\label{sec:singlecrossing}

The conditions in Eqs.~(\ref{consistency-1}) and
(\ref{consistency-2}) allow us to understand some of the stability
features analytically. For example, for the integral in
Eq.~(\ref{consistency-2}) to vanish, the integrand has to be
positive in some parts and negative in the other. Thus, an
instability requires the spectrum $g_\omega$ to have a
zero-crossing. The existence of such an instability also requires the
spectrum  to cross from negative to positive values \cite{Dasgupta:2009mg}.

\begin{figure}[h]
\begin{center}
\includegraphics[width=0.8\columnwidth]{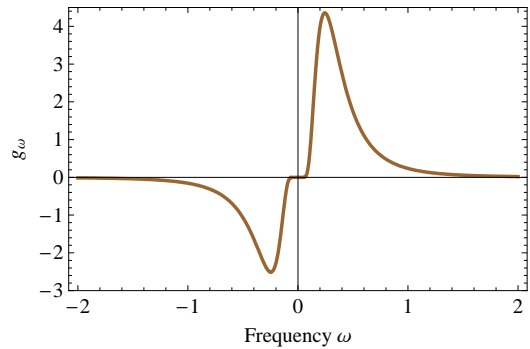}
\caption{Fermi-Dirac spectrum for $\nu_e$ and $\bar\nu_e$,
with $\Delta m^2 =1$, and $T=1$ for both species.
A small degeneracy parameter $\eta=0.28$ provides an excess flux
of $\nu_e$ over $\bar\nu_e$, which corresponds to $\epsilon=0.666$.
\label{fig:FDspectrum}}
\end{center}
\end{figure}

During the accretion phase of SN evolution, powerful $\nu_e$ and
$\bar\nu_e$ fluxes are emitted, with a much weaker flux of other
flavors. If we model this situation by a pure Fermi-Dirac spectrum
of $e$-flavored neutrinos in the inverted hierarchy, the spectrum is
positive for $\nu_e$ and negative for $\bar\nu_e$, providing for a
single-crossed spectrum (Fig.~\ref{fig:FDspectrum}). Therefore,
instabilities are a generic feature of the neutrino flux streaming
from a SN core.

Note that in the limit $\kappa \to 0$, the integrand in
Eq.~(\ref{consistency-2}) becomes $g_\omega \pi \delta(\omega-\gamma)$.
Therefore, the consistency condition implies $g(\gamma)=0$.
Thus in this limit, the instability starts near the zero crossing.
Also, the r.h.s.\ of Eq.~(\ref{consistency-1}) gets a large contribution
from the modes $\omega \approx \gamma$, implying a small $\mu$.
Therefore in the single-angle approximation, at small neutrino density
one generically gets a narrow instability centered at a positive crossing.

\subsection{Normal vs. inverted hierarchy}
\label{hierarchy-single}

Recall that all our results have been obtained using inverted
hierarchy. Going to normal hierarchy corresponds to changing
Eq.~(\ref{eom-single}) to
\begin{equation}
\I\partial_r \tilde{S}_{\omega}= - \omega\, \tilde{S}_{\omega}
-2 u_0 \mu \int d\omega'\,g_{\omega'}\, \tilde{S}_{\omega'}\,.
\label{eom-single-h}
\end{equation}
In terms of the solution $S_{\omega}$ of Eq.~(\ref{eom-single}), the
solution of this equation is given by
\begin{eqnarray}
\tilde{S}_{\omega}(\mu,g_\omega) & = & S^*_{\omega}(\mu, -g_\omega)
= S^*_{\omega}(-\mu, g_\omega) \; .
\end{eqnarray}
Since  $S$ and $S^*$ should have the same stability behavior, this
implies that the stability conditions for normal hierarchy are the
same as those for the inverted hierarchy with a change in the sign
of $g_\omega$.

Formally we may implement this change of sign by keeping the same
$g_\omega$ and $\mu\to-\mu$. In this sense one can show the
solutions $\kappa(\mu)$ and $\gamma(\mu)$ on the same plot for both
hierarchies by extending it to negative values of $\mu$.

\subsection{Multiple spectral crossings}

A supernova emits neutrinos of all flavors, although with different
fluxes and spectra. As $g_\omega$ is the difference spectrum between
$\nu_e$ and $\nu_x$, usually there will be additional spectral
crossings other than the one at $\omega=0$. In the limit $\mu\to0$
and concomitant $\kappa\to 0$, there will be a solution $Q_\omega$
centered on every positive crossing as already stressed in
Ref.~\cite{Dasgupta:2009mg}. For larger $\mu$ there can be fewer
solutions, but if there are several, they can coexist even when the
different eigenfunctions are not well separated.

In the single-angle case, our problem is equivalent to the reduced
pairing Hamiltonian which is at the core of the BCS theory of
superconductivity~\cite{Pehlivan:2011hp}. In this context, the same
stability problem was recently investigated and it was shown how
general analytic solutions can be
constructed~\cite{Yuzbashyan:2008}. In other words, explicit
large-amplitude solutions were constructed that correspond to
several coexisting solutions with different $\Omega$-values. The
single-$\Omega$ large-amplitude solution of
Ref.~\cite{Dasgupta:2009mg} is the simplest case of the class of
multiple ``normal soliton solutions.''

\section{Box-spectra examples}
\label{sec:single-ex}

In order to understand the behavior of the stability criteria it is
useful to study explicit examples of single and multi-crossed
spectra that can be solved analytically. The simplest approach is to
represent a spectrum such as Fig.~\ref{fig:FDspectrum} in the
single-energy approximation by one $\delta$ function for one average
$\omega$ (neutrinos) and one for negative $\omega$ (antineutrinos),
or more such spikes to represent a multi-crossed spectrum. One then
finds simple polynomial equations for $\Omega$ with coefficients
that depend on the frequencies and heights of the spikes.

We here go one step further and represent the spectra by ``boxes''
of unit height, i.e.\ $g_\omega$ only takes the values $0,\pm 1$.
This makes the integral in Eq.~(\ref{consistency-single}) once more
analytically calculable and $\Omega$ becomes the root of a
polynomial whose degree depends on the number of boxes. As will be
seen, the simplified box spectra already bring out some important
features of the stability of the realistic SN spectra.

Scaling the height of all the boxes by a factor $\alpha$ corresponds to
scaling the spectrum as $g_\omega \to \alpha g_\omega$. The results for
this may be obtained by simply scaling $\mu \to \alpha \mu$. For the sake
of simplicity, in the analytic single-angle arguments we use $u_0 =
1/2$. The results for any $u_0$ may be obtained by the scaling $\mu
\to 2 u_0 \mu$. The numerical results are given for various sample
$u_0$ values.

\subsection{Two boxes}
\label{two-box}

\begin{figure}
\begin{center}
\hspace{-0.2in}
\includegraphics[width=0.8\columnwidth]{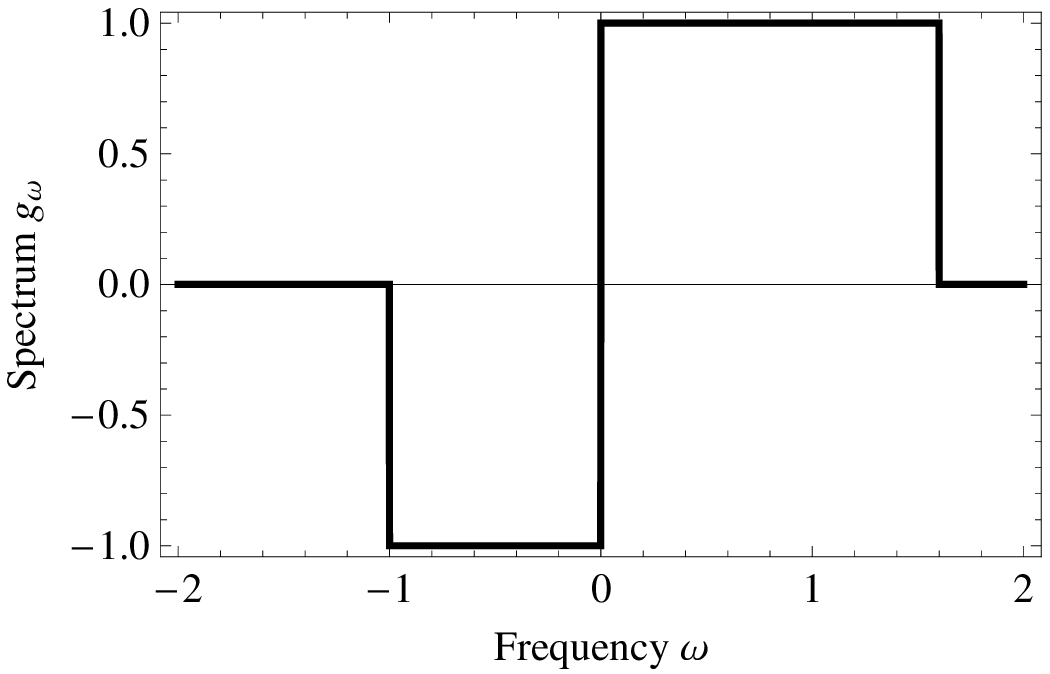}\\
\vspace{0.2in}
\includegraphics[width=0.8\columnwidth]{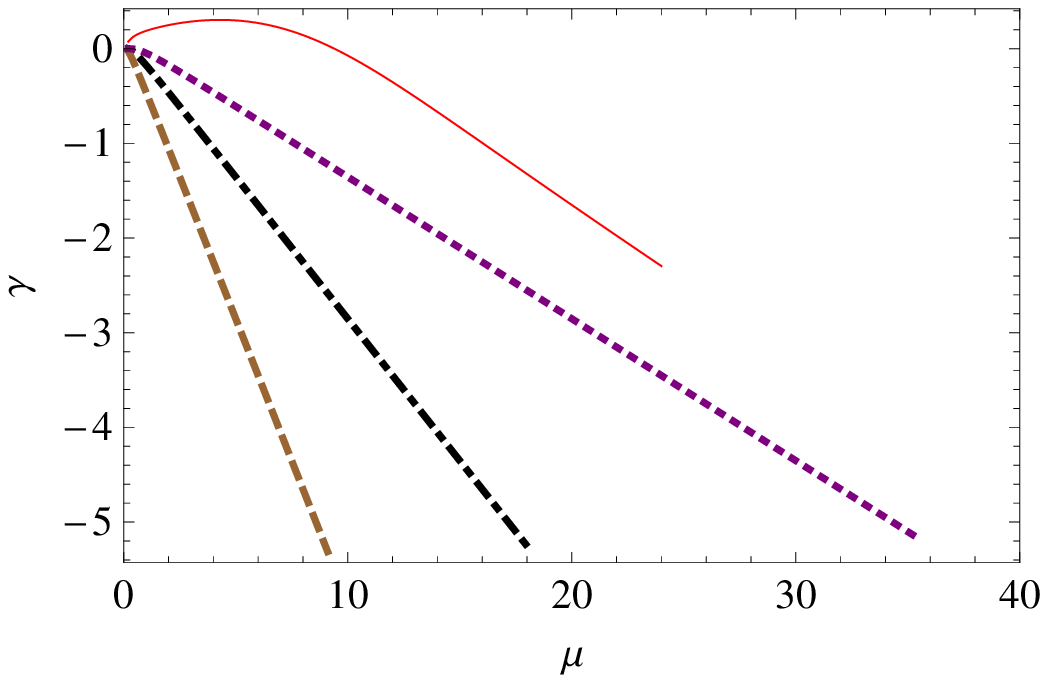}\\
\vspace{0.1in}
\includegraphics[width=0.8\columnwidth]{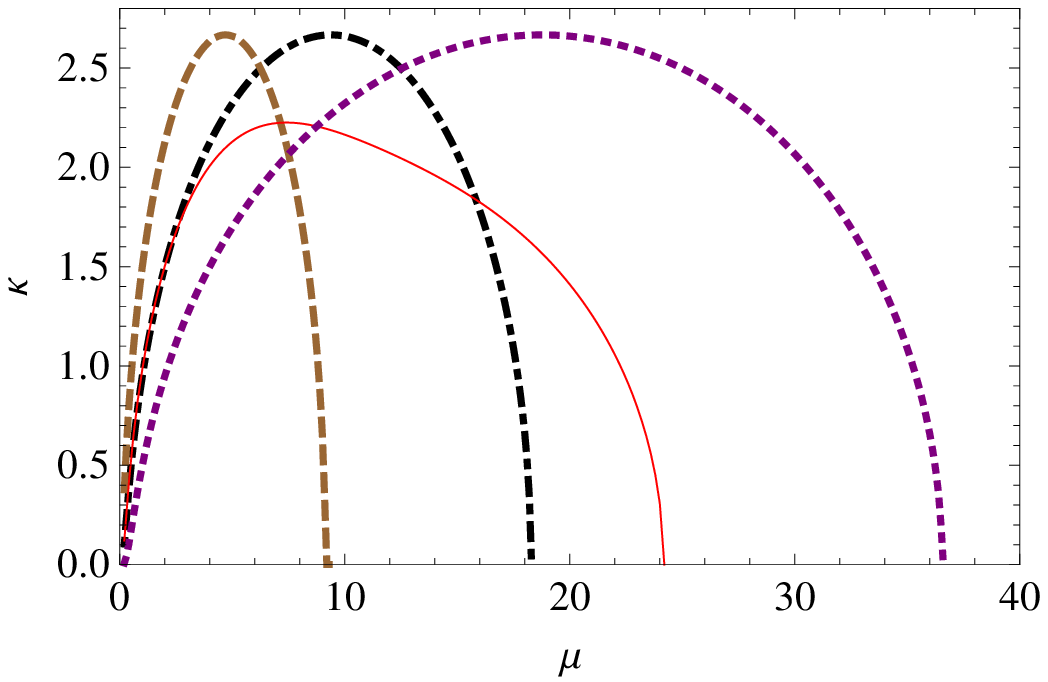}
\caption {Two-box spectrum with $a=1$ and $b=1.6$.
Lower panels: $\gamma(\mu)$ and $\kappa(\mu)$, where broken
curves are single angle with $u_0=1$, $1/2$ and $1/4$ (left to right).
Thin solid (red) line: Multi-angle with uniform
distribution $0\leq u\leq1$.
Solutions for $\gamma(\mu)$ are only shown where $\kappa(\mu)\not=0$.}
\label{fig:2box-single}
\end{center}
\end{figure}

Our first example is a schematic representation of the Fermi-Dirac
spectrum of Fig.~\ref{fig:FDspectrum}. We define our two-box
spectrum~as
\begin{equation}
 g_{\omega}=\left\{\begin{matrix}
 -1 && && -a < \omega< 0 \\ +1 && && 0<\omega<b \\
\end{matrix} \right.
\end{equation}
as shown in Fig.~\ref{fig:2box-single}. We assume that all frequencies
are normalized to some common frequency scale.
The consistency condition in Eq.~(\ref{consistency-single}) yields
\beq
 \frac{\Omega^2}{(\Omega+a)(\Omega-b)}=\eta \; ,
\eeq
where $\eta \equiv e^{-1/\mu}$. This corresponds to the
quadratic equation
\beq
(1-\eta)\Omega^2+(b\eta-a\eta)\Omega+a b \eta=0  \; ,
\label{2beq}
\eeq
from which the stability can be analyzed analytically.
Note that $0 < \eta < 1$, while the limits
$\mu \to \infty$ and $\mu \to 0$ correspond to
$\eta =1$ and $\eta=0$, respectively.

The roots of Eq.~(\ref{2beq}) are
\begin{equation}
{\sf\Omega}=\frac{-(b-a)\eta\pm\sqrt{(b-a)^2\eta^2-4a b\eta (1-\eta)}}
{2(1-\eta)}\,.
\end{equation}
We have two complex conjugate solutions when the argument of
the square-root is negative, i.e.\ for
\begin{equation}
0 < \eta< \eta_{\rm sync} \equiv \frac{4ab}{(b+a)^2}\,.
\end{equation}
Here $\eta_{\rm sync}$ is the ``synchronization strength,'' i.e.\ for a
larger interaction strength the system is stuck in a stable
position. If $0<\eta<\eta_{\rm sync}$, we find
\begin{eqnarray}
\gamma&=&-\frac{(b-a)\eta}{2(1-\eta)}\,,
\nonumber\\
\kappa&=&\pm\frac{\sqrt{4ab\eta(1-\eta)-(b-a)^2\eta^2}}{2(1-\eta)}\,.
\end{eqnarray}
These solutions are shown in Fig.~\ref{fig:2box-single} for a spectrum with
$a=1$, $b=1.6$, and different
values of the single emission angle $u_0$.

For $\eta>\eta_{\rm sync}$ the solutions for $\Omega$ are two different
real roots so that $\kappa=0$.
At the synchronization strength $\eta_{\rm sync}$, we find
\begin{equation}
\gamma_{\rm sync}=-\frac{2ab}{(b-a)}\,.
\end{equation}
In our figures we only show the solutions with nonvanishing
$\kappa$, since when $\kappa=0$, the complex solutions $\gamma \pm
i\kappa$ do not exist and there are no instabilities.
Therefore, the $\gamma(\mu)$ curves are only shown where
$\kappa(\mu)\not=0$.

At the vanishing interaction strength $\mu=0$, we have
$\gamma=\kappa=0$. Had we put a spectral gap between the two boxes,
the lower-$\mu$ cutoff point would be at a nonzero interaction
strength $\mu >0$.

For the completely antisymmetric two-box spectrum ($a=b$), Eq.~(\ref{2beq})
reduces to
\begin{equation}
\Omega^2=-\frac{a b \eta}{1-\eta} \; ,
\label{2blo}
\end{equation}
which always has purely imaginary solutions. Thus, there is
no synchronized behavior for such a spectrum, the flavor conversions
take place at arbitrarily high values of $\mu$. We will not further consider
this special case that requires a fine-tuned spectrum. Moreover,
a SN core always produces
an excess flux of $\nu_e$ over $\bar\nu_e$ due to deleptonization.

The spectrum shown in Fig.~\ref{fig:2box-single} has a positive
zero-crossing. The case of a negative zero-crossing may be studied
by multiplying the spectrum $g_\omega$ by a factor of $-1$, i.e.\
$g_\omega \to -g_\omega$.
The consistency conditions then stay the same, with the change
$\mu \to -\mu$, or $\eta \to 1/\eta$.
Equation~(\ref{2beq}) then becomes
\beq
(1-\eta) \Omega^2 - (b-a) \Omega - a b =0 \; ,
\eeq
which always has real roots. Thus, the spectrum is stable as anticipated
from our earlier arguments.

To summarize, for the two-box spectrum in the single-angle approximation,
the spectrum with a positive zero-crossing is always stable in the
normal hierarchy. In the inverted hierarchy
(i) the spectrum with a negative zero-crossing is always stable,
(ii) a completely antisymmetric spectrum with a positive zero-crossing
is unstable at any value of $\mu$,
(iii) an asymmetric spectrum with a positive zero-crossing is stable
above a threshold value of the interaction strength and unstable if
the interaction strength is smaller.

\subsection{Three boxes}
\label{three-box}

The natural next case is a spectrum with two crossings, of which only
one can be positive, and so we expect at most one unstable solution
for the eigenvalue equation. However, it was recognized earlier that
such a spectrum shows additional features in that $\kappa(\mu)$ need
not cut off at large $\mu$ \cite{Raffelt:2008hr}.

\begin{figure}[h]
\begin{center}
\hspace{-0.05in}\includegraphics[width=0.8\columnwidth]{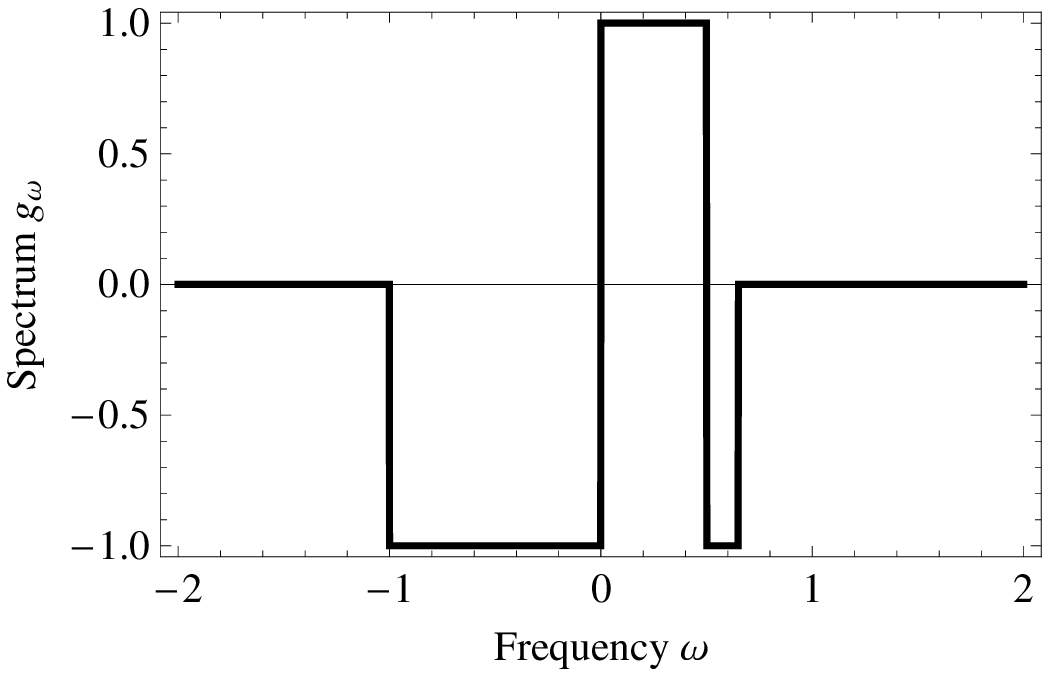}\\
\vspace{0.2in}
\includegraphics[width=0.8\columnwidth]{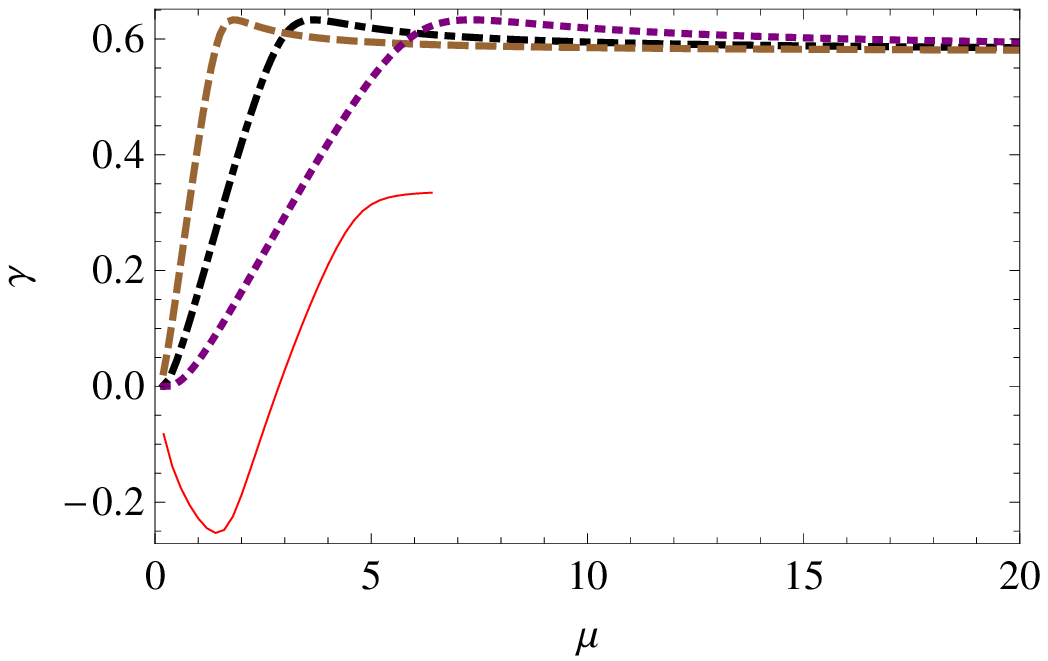}\\
\vspace{0.1in}
\includegraphics[width=0.8\columnwidth]{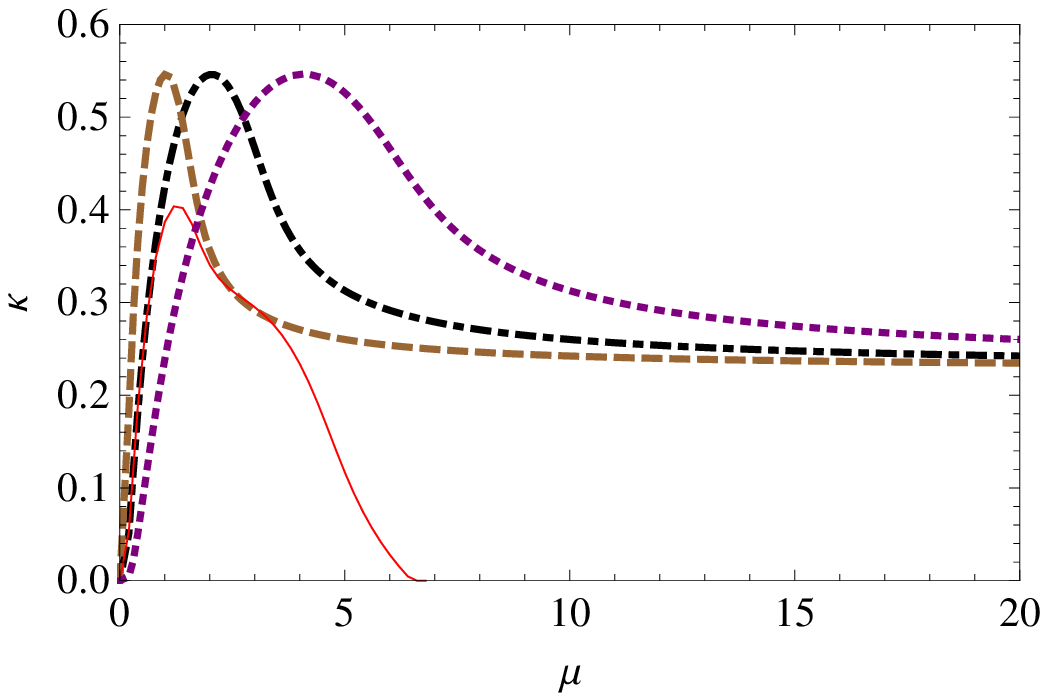}
\caption {Three-box spectrum with $a=1, b=0.5, c=0.6$. {\it Lower
panels}: $\gamma(\mu)$ and $\kappa(\mu)$, where broken
curves are single angle with $u_0=1$, $1/2$ and $1/4$ (left to right).
Thin solid (red) line: Multi-angle with uniform
distribution $0\leq u\leq1$.
Solutions for $\gamma(\mu)$ are only shown where $\kappa(\mu)\not=0$.}
\label{fig:3box-single}
\end{center}
\end{figure}

We represent this case by three adjacent equal-height boxes of the form
(Fig.~\ref{fig:3box-single})
\begin{equation}
 g_{\omega}=\left\{\begin{matrix}
 -1 && && -a<\omega<0 \\ +1 && && 0<\omega<b \\ -1 && && b<\omega<c
\end{matrix} \right. \;.
\end{equation}
The consistency condition in Eq.~(\ref{consistency-single})
leads to the cubic equation
\begin{eqnarray}
(1-\eta)\Omega^3-(c-2 b\eta+a\eta)\Omega^2 & & \nonumber \\
 -(b^2 - 2 a b)\eta \Omega-a b^2 \eta& =0\;.&
\label{3beq}
\end{eqnarray}
It can have three real roots, or a single real root and a pair of
complex conjugate roots. The latter case corresponds to instability.

In order to study the stability at large $\mu$, we look at the limit
$\eta = 1$. The cubic term in Eq.~(\ref{3beq}) drops out,
leaving us with the quadratic equation
\begin{equation}
 (a-2 b +c)\Omega^2+(b^2-2 a b)\Omega+a b^2=0 \; .
\label{3b2}
\end{equation}
When the total lepton number is not zero ($2 b -a -c\neq0$)
the solutions are
\begin{equation}
 \Omega=\frac{2 a b -b^2\pm \sqrt{b(4 ab +b^2-4 a c)}}{2(a-2 b +c)} \; .
\end{equation}
If $4 ab +b^2-4 a c>0$ then both roots are real, implying that
the system is stable.
We then get the same cutoff behavior as in the two-box
case, similar to the one shown in Fig.~\ref{fig:2box-single}.

On the other hand, if $4 ab +b^2-4 a c<0$, and writing
$\Omega = \gamma \pm i \kappa$, we find
\beq
\gamma=\frac{2 a b -b^2}{2(a-2 b +c)} \;, \quad
\kappa=\frac{\sqrt{b(4 ac -b^2-4 a b)}}{2(a-2 b +c)} \; .
\eeq
Thus $\kappa$ reaches a non-zero asymptotic saturation value that is
independent of $\mu$. Likewise, $\gamma$ approaches a $\mu$-independent
asymptotic value, implying that the eigenfunction $Q_\omega$ remains roughly
centered on the nonvanishing spectral range.
The system is thus unstable even at large $\mu$ values.
This saturation behavior, shown in Fig.~\ref{fig:3box-single},
was absent in the two-box scenario. This would allow flavor
conversions to start deep inside the SN core.

The spectrum in Fig.~\ref{fig:3box-single} has its central box
positive. The corresponding spectrum with the central box negative
can be treated in the same manner, with the replacement $g_\omega
\to -g_\omega$, which corresponds to $\mu \to -\mu$ or $\eta \to
1/\eta$. In the $\eta=1$ limit (large neutrino density) the behavior
is then identical with the one described above.

When the total lepton number vanishes, Eq.~(\ref{3beq}) reduces to
\begin{equation}
(1-\eta)\Omega^3-(2 b-a)(1-\eta)\Omega^2-(b^2 - 2 a b)\eta \Omega-a b^2 \eta
=0 \; .
\end{equation}
One can examine if it has three roots of the form $\Omega_1,
\Omega_2 = \gamma + i\kappa, \Omega_3 = \gamma - i\kappa$ in the limit
$\eta = 1$. Clearly in this limit, $\Omega_1 = ab/(2a-b)$ is a real finite
root. Then the sum of the roots is $\Omega_1 + 2 \gamma = 2b-a$, a finite
number, implyting that $\gamma$ is finite. On the other hand,
the product of the roots, $\Omega_1 (\gamma^2 -\kappa^2)
= - a b^2 \eta /(1-\eta)$ is a large negative number. This is possible
only if $\kappa$ is a large number, and $\Omega_1 >0$, i.e.
$2 a -b >0$.
Thus, the stability in this case is determined by
the sign of $(2a -b)$. If this is negative, then there is a cutoff
behavior, i.e.\ the system is stable in the $\eta = 1$ limit.
If this is positive, then the system is unstable at arbitrarily large $\mu$.
For the inverted spectrum $g_\omega \to -g_\omega$,
the condition for stability is reversed.

The three-box spectra thus show a cutoff or saturation
behavior, depending on the details of the spectrum. The saturation
behavior cannot occur for two boxes and is a new feature.
The three-box spectra thus open up the possibility
that for certain combinations of spectra and hierarchy, the neutrino
ensemble is never stable. We will see, however, that this behavior is qualitatively
modified by multi-angle effects.

\subsection{Four boxes}
\label{four-box}

\begin{figure}
\begin{center}
\includegraphics[width=0.8\columnwidth]{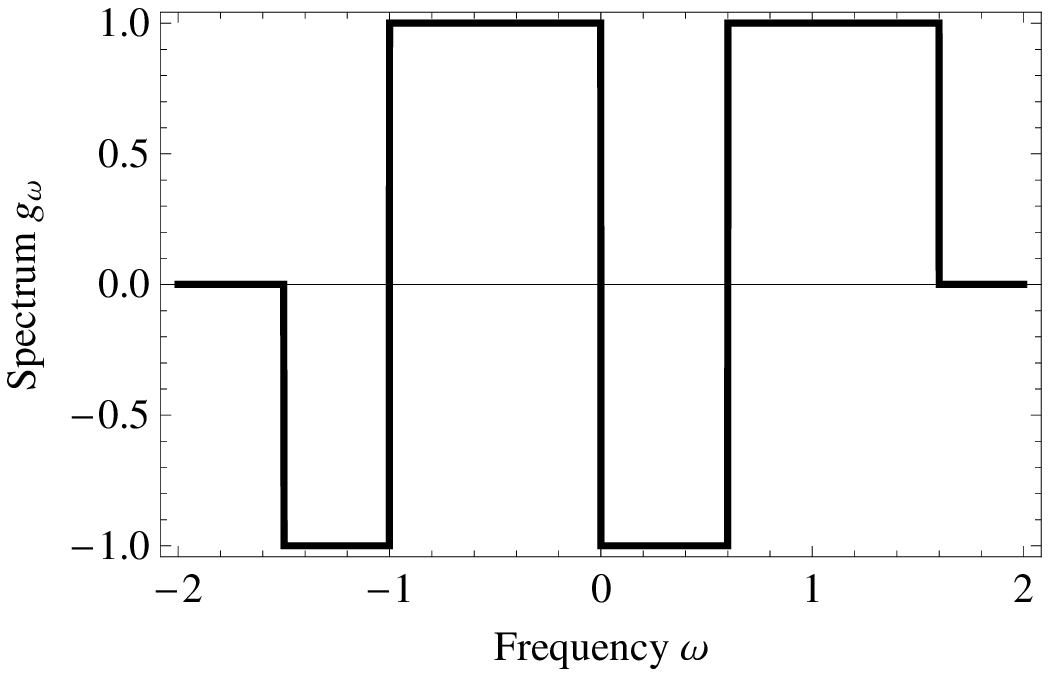}\\
\vspace{0.2in}
\includegraphics[width=0.8\columnwidth]{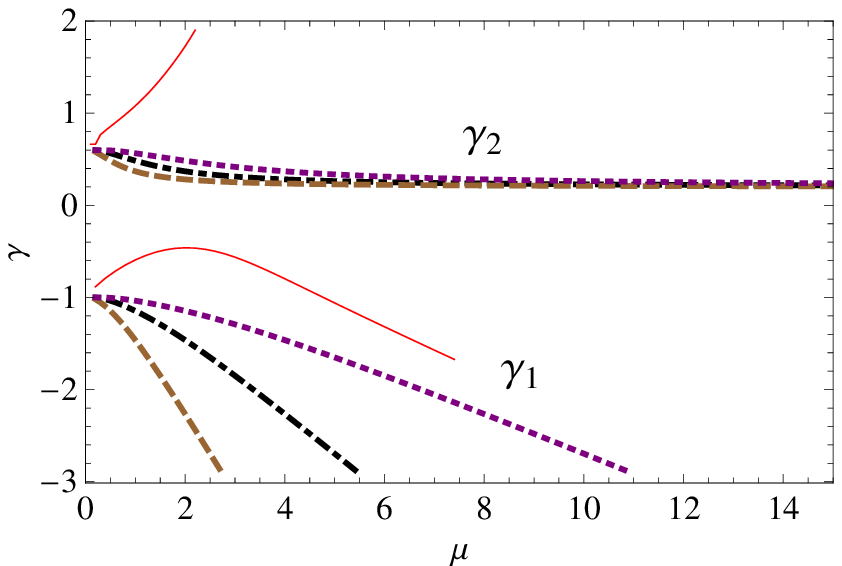}\\
\vspace{0.1in}
\includegraphics[width=0.8\columnwidth]{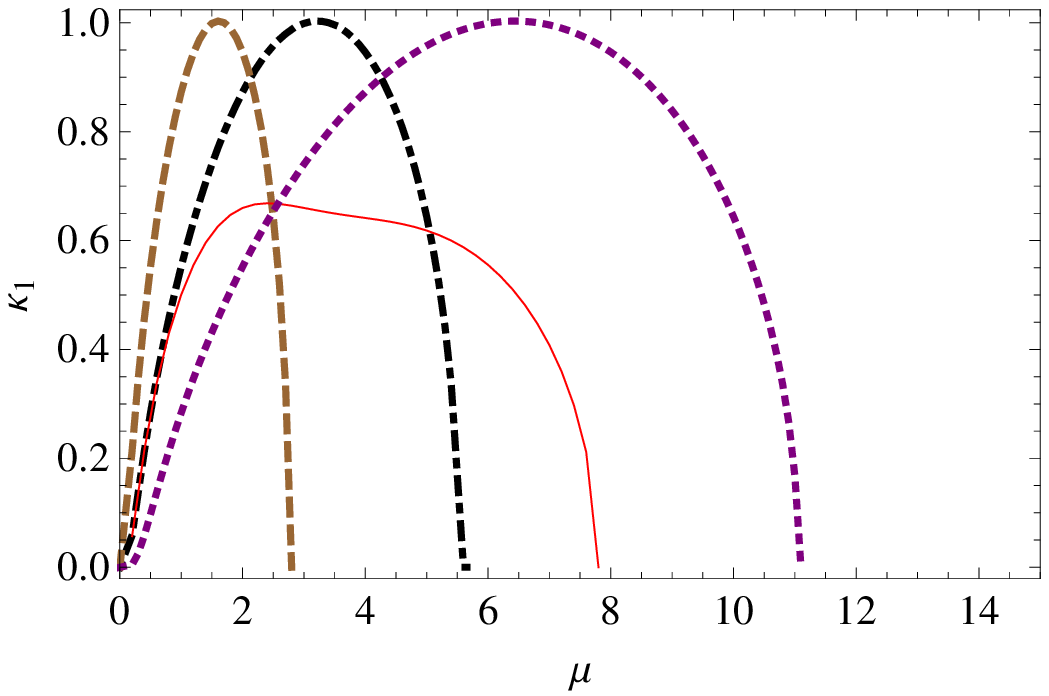} \\
\vspace{0.1in}
\includegraphics[width=0.8\columnwidth]{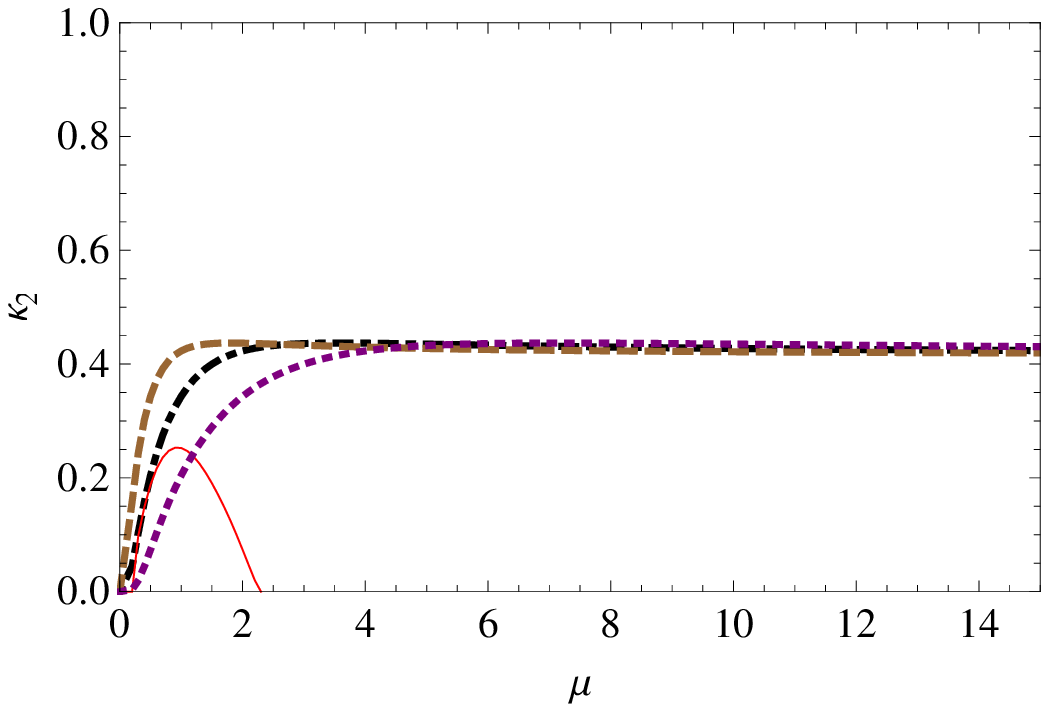}
\caption {Four-box spectrum with $a=1.5$, $b=1.0$, $c=0.6$ and $d=1.6$.
{\it Lower
panels}: $\gamma(\mu)$ and $\kappa(\mu)$, where the solution with larger
$\gamma$ is termed number 1. Broken
curves are single angle with $u_0=1$, $1/2$ and $1/4$ (left to right).
Thin solid (red) line: Multi-angle with uniform
distribution $0\leq u\leq1$.
Solutions for $\gamma(\mu)$ are only shown where $\kappa(\mu)\not=0$.}
\label{fig:4box-single}
\end{center}
\end{figure}

A four-box spectrum comes closest to representing a realistic SN
spectrum which typically has three zero crossings as explained earlier.
Representing it by four adjacent boxes, we define the spectrum as
\begin{equation}
 g_{\omega}=\left\{\begin{matrix}
- 1 && && -a<\omega<-b \\ +1 && && -b<\omega<0 \\
- 1 && && 0<\omega<c \\ +1 && && c<\omega<d
\end{matrix}
\right. \; .
\end{equation}
The self consistency condition is
\begin{equation}
 \frac{(\Omega+b)^2(\Omega-c)^2}{(\Omega+a)\Omega^2(\Omega-d)}=\eta \;
\end{equation}
and can be written as
\begin{eqnarray}
  (1-\eta)\Omega^4+(2 b - 2 c - a \eta+d \eta)\Omega^3 && \nonumber
\\ +(b^2-4 b c +c^2 +a d \eta)\Omega^2  \nonumber \\
+(2 b c^2-2 b^2 c)\Omega +b^2c^2&=&0 \; .
\label{4beq}
 \end{eqnarray}
This quartic equation can have zero, one or two pairs of complex solutions, so this
is the first explicit case with the possibility of two simultaneous
unstable solutions.

We are interested in the high-density behavior where
$\eta = 1$. In this limit,
the quartic equation reduces to a cubic equation with real coefficients.
Hence we are guaranteed at least one real root and
we can have at most one pair of complex conjugate solutions, i.e.\ at most
one solution for $\kappa^2$.
Its existence is determined by
the value of the discriminant
\begin{eqnarray}
\Delta&=&b^2 c^2 \bigl[ -27 b^2 c^2 (a-2 b+2 c-d)^2\nonumber \\
&+&32 b (b-c)^3 c (-a+2 b-2 c+d)\nonumber \\
&-&36 b (b-c) c (-a+2 b-2 c+d) (b^2-4 b c+c^2+a d)\nonumber \\
&+&4 (b-c)^2 (b^2-4 b c+c^2+a d)^2\nonumber \\
&-&4 (b^2-4 b c+c^2+a d)^3\bigr] \; .
\end{eqnarray}
If $\Delta>0$, all the roots are real and the system is stable.

A typical scenario is shown in Fig.~\ref{fig:4box-single}.
At low $\mu$ values, there are two instabilities, one of which
shows a cutoff behavior, i.e.\ it vanishes for $\mu$ greater
than a certain value.
The other instability, with $\gamma \approx 0$, survives
for arbitrarily large values of $\mu$, with $\kappa$ showing a
saturation behavior as in the 3-box case.
Under the transformation $g_\omega \to -g_\omega$, $\mu \to -\mu$,
$\eta \to 1/\eta$, the instability condition at $\eta=1$ does not
change, so that at large neutrino densities the saturation solution exists
in either hierarchy.

The four-box spectra can give rise to two instabilities at the same time,
or only one, or none at all, depending on the spectral details
and strength of $\mu$.
At most one of them can show saturation behavior at large density, which
then exists for both hierarchies.

\section{Spectra with tails}
\label{sec:nosleep}

The ultimate goal of our investigation in the context of SN physics
is to understand in which regions of the star the neutrino stream
may show self-induced instabilities. In agreement with the
previous literature we have found that in some cases, and notably in
the generic two-box example, there is a value $\mu_{\rm sync}$ such
that for $\mu>\mu_{\rm sync}$ the system is stable. Upon closer
scrutiny, however, one finds that the stable regime does not exist
for a realistic spectrum. The main difference between the
Fermi-Dirac spectrum of Fig.~\ref{fig:FDspectrum} and the two-box
spectrum of Fig.~\ref{fig:2box-single} is that the former has
nonvanishing tails for large $|\omega|$, corresponding to the
infrared part of the spectrum. In other words, $g_\omega\not=0$
everywhere except at the spectral crossing point. Given a positive
spectral crossing, the only solutions with $\kappa\to0$ can be those
centered on the spectral crossing and corresponding to $\mu\to0$.
Purely real solutions for $\Omega$ then do not exist for large $\mu$
and the system is always unstable.

\begin{figure}
\begin{center}
\includegraphics[width=0.8\columnwidth]{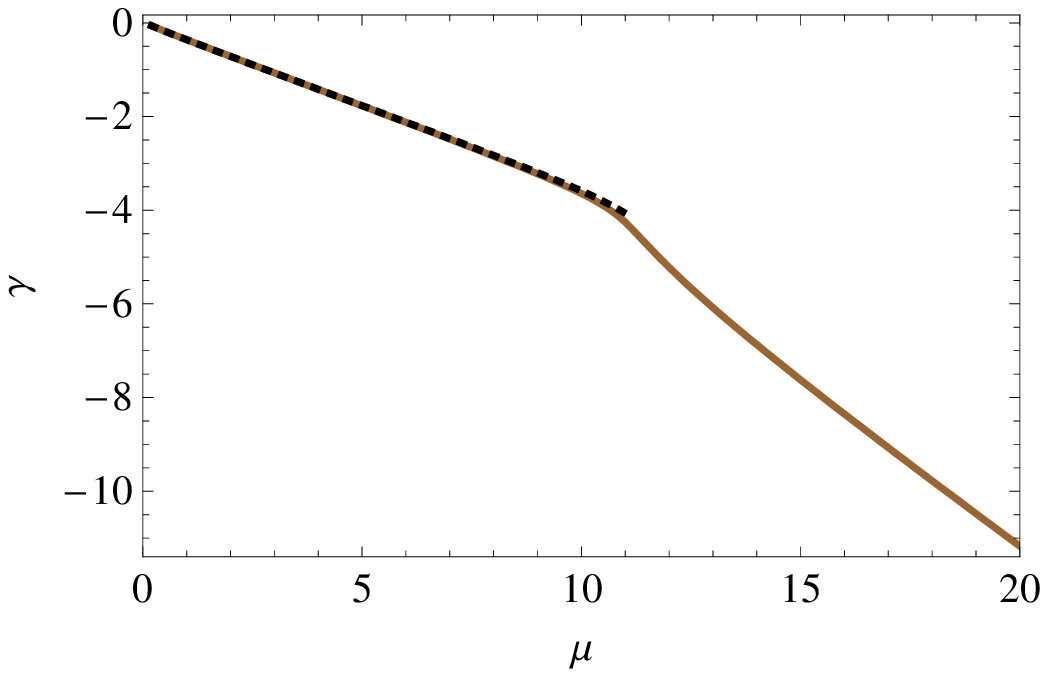}
\includegraphics[width=0.8\columnwidth]{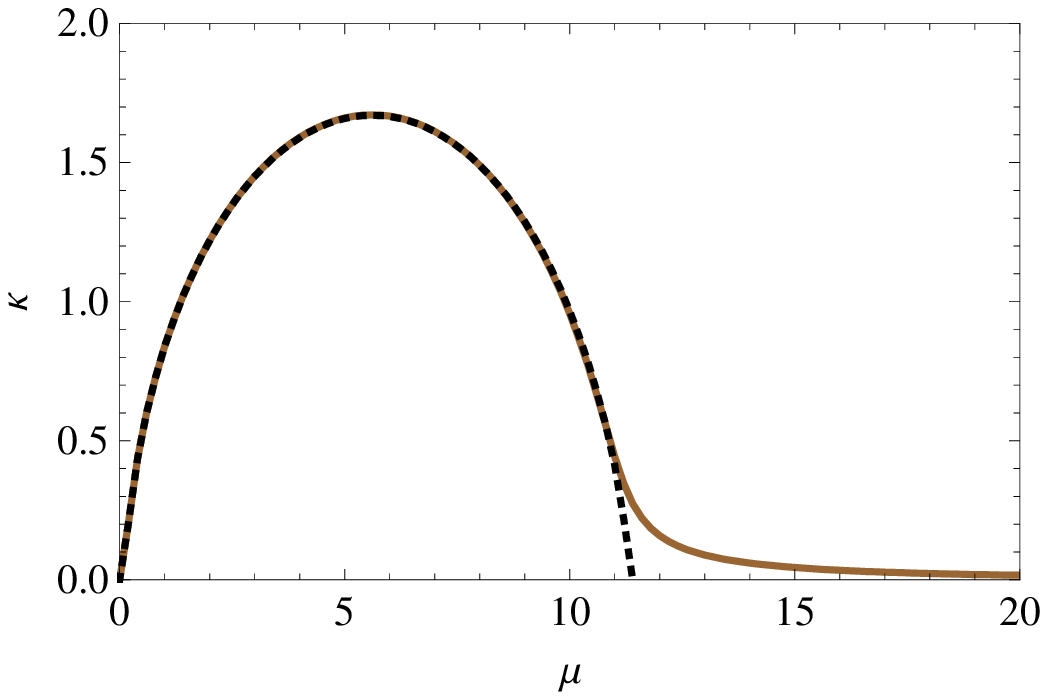}
\includegraphics[width=0.8\columnwidth]{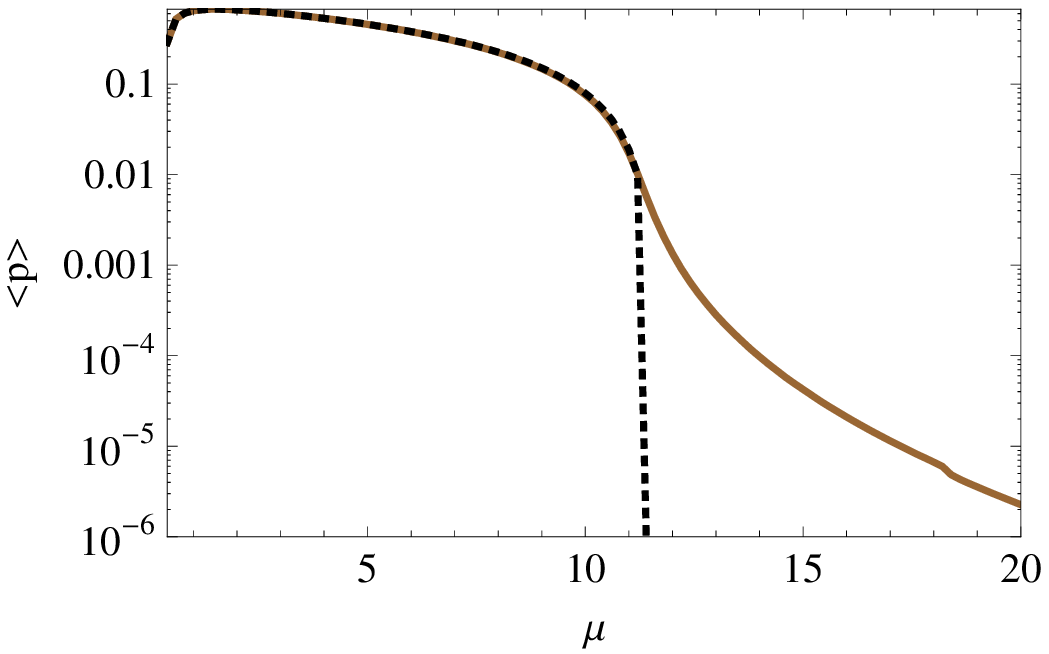}
\caption{Eigenvalue solution for the Fermi-Dirac spectrum of
Fig.~\ref{fig:FDspectrum}, shown by the brown (solid) line.
As a dotted (black) line we show the same result if the Fermi-Dirac spectrum
is cut off for $|\omega|>4$.
Bottom panel: $\langle p \rangle$. \label{fig:FDsolution}}
\end{center}
\end{figure}

One can easily illustrate this point by a numerical solution of the
eigenvalue equation for the Fermi-Dirac spectrum of
Fig.~\ref{fig:FDspectrum}. We show $\gamma(\mu)$ and $\kappa(\mu)$
in Fig.~\ref{fig:FDsolution}, assuming single angle with $u_0=1/2$.
The result is qualitatively similar to the two-box example, except
that at large $\mu$ the curve $\kappa(\mu)$ has a tail.

In a SN core, the system as a function of radius sweeps from very
large $\mu$-values (of order $10^6$ in our normalization) to zero,
but the system is nowhere stable. Still, the usual onset of bipolar
oscillations happens approximately at the $\mu$-value around the
would-be synchronization value where the ``hump'' and the ``tail''
of the $\kappa(\mu)$ curve join, in Fig.~\ref{fig:FDsolution} around
$\mu\sim10$--11, as confirmed in numerous numerical SN examples.

The explanation is two-fold. At large $\mu$ the system is always
unstable, but $\kappa(\mu)$ is very small and the exponential growth
will not get anywhere on the available time scale (or rather, radial
distance scale), in particular as the center of the resonance,
$\gamma(\mu)$, is constantly shifting as a function of the
decreasing $\mu$. Second, the center of the resonance is far away in
the spectral tail (the infrared energy modes), so only a narrow
range of infrared modes is affected. It deserves mention that
``small $\kappa$'' simultaneously means two things: The
exponential growth is slow and the resonance width is narrow.

Evidently, then, a stability analysis alone is not enough to
understand the consequences in a realistic SN. One needs a more
dynamical approach that takes account of the available time scale as
$\mu$ sweeps from large to small values, and one needs to consider
the range of modes that are actually affected by the resonance.

One rough way to approach the latter point is to consider the
maximum total flavor conversion that can be achieved by a solution
$\Omega$ on a given spectrum, or equivalently, the average
conversion probability over the entire spectrum when the flavor
pendulum has moved from the inverted to the normal position. From
the exact large-amplitude instantaneous solution \cite{Dasgupta:2009mg}
one finds that the maximum
conversion probability of a mode of frequency $\omega$ is
\begin{equation}
p_\omega=\frac{\kappa^2}{(\omega-\gamma)^2+\kappa^2}\,.
\end{equation}
Averaged over all modes this is
\begin{equation}
\langle p \rangle =\frac{\int_{-\infty}^{+\infty}d\omega\,p_\omega|g_\omega|}
{\int_{-\infty}^{+\infty}d\omega\,|g_\omega|}\,.
\end{equation}
We can speak of an effective cutoff behavior if $\langle p \rangle$
as a function of $\mu$ quickly drops to very small values in a
narrow range of $\mu$, while being appreciably large at smaller
$\mu$. In the bottom panel of Fig.~\ref{fig:FDsolution} we show
$\langle p \rangle$ for the Fermi-Dirac example. Indeed,
for $\mu\sim11$, this measure drops by orders of magnitude and we
would have an effective cutoff---above this transition range, only a
minimal range of infrared modes experiences self-induced conversion.

Another approach is to artificially cut off the infrared part of the
spectrum. The curve $\gamma(\mu)$ in Fig.~\ref{fig:FDsolution} has a
distinct kink exactly in the region where the the hump and tail of
the $\kappa(\mu)$ curve join. If we cut off the Fermi-Dirac spectrum
and set it to 0 for $|\omega|>4$, the $\kappa(\mu)$ curve follows
the original one almost exactly, except that the tail is cut off at
$\mu_{\rm sync}\sim 11$ and the system becomes perfectly
stable for larger $\mu$ values (Fig.~\ref{fig:FDsolution}).
Therefore, the simple picture of a cutoff behavior is actually very
good except for a minor infrared correction and our study of box
spectra indeed provides a useful picture of the instability
behavior.

The saturation solutions that can arise for spectra with two or more
crossings (three or more boxes) are very different. The saturation
solutions have a $\kappa$ that is comparable to the width of the
spectrum and remain centered on the main part of the spectrum.
Therefore, such an instability cannot be removed by truncating the
infrared part of the spectrum. These instabilities are only tamed by
multi-angle effects.

\section{Multi-angle stability analysis}
\label{sec:multi}

\subsection{The consistency conditions}
\label{eigen:multi}

The linearized stability analysis in the single-angle case
shed new light on the self-consistency conditions as originating
from a simple eigenvalue equation, although the conditions
themselves had been found previously together with the full solution
independently of the small-amplitude expansion. However, the
single-angle approximation is not justified in the SN context except
that numerically single-angle simulations and multi-angle simulation
sometimes, but not always, yield similar results. Therefore, based
on the eigenvalue equation~(\ref{fourier-eom}), we now extend the
linearized analysis to the multi-angle case. Besides the
small-amplitude expansion, this equation also uses the
large-distance approximation where it is assumed that the angular
divergence of the neutrino radiation is small.

The crucial step is to realize that the r.h.s.\ of
Eq.~(\ref{fourier-eom}) is of the form $A+B u$ where $A$ and $B$ are
expressions that do not depend on either $\omega$ or $u$. So we are
led to the ansatz for the form of the eigenfunction
\begin{equation}
Q_{\omega,u}=\frac{a+b\,u}{\omega+u\bar\lambda-\Omega}\,,
\end{equation}
where $a$ and $b$ are complex numbers. Inserting this ansatz
provides
\begin{equation}
a+bu=\mu
\int du'\,d\omega'\,g_{\omega',u'}\,
\frac{(u+u')(a+b\,u')}{\omega'+u'\bar\lambda-\Omega}\,.
\end{equation}
To understand better the structure of this equation, we define the
integrals
\begin{equation}
I_n=\int du\,d\omega\,g_{\omega,u}\,
\frac{u^n}{\omega+u\bar\lambda-\Omega}\,.
\label{In-def}
\end{equation}
Then our eigenvalue equation becomes
\begin{equation}
a+bu=\mu\Bigl[(aI_1+bI_2)+(aI_0+bI_1)u\Bigr]\,.
\end{equation}
If this is supposed to be true for every $u$ we need to match the
coefficients of the linear $u$ polynomial on both sides separately.
We can then write this in matrix form
\begin{equation}
\mu^{-1}
\begin{pmatrix} a\\ b\end{pmatrix}=
\begin{pmatrix} I_1&I_2\\ I_0&I_1\end{pmatrix}
\begin{pmatrix} a\\ b\end{pmatrix}\,.
\end{equation}
This has the form of an eigenvalue equation for a $2\times 2$ matrix.
This equation has nontrivial solutions if
\begin{equation}
{\rm det}\left[\begin{pmatrix} I_1&I_2\\ I_0&I_1\end{pmatrix}-\mu^{-1}\right]=0
\end{equation}
or explicitly
\begin{equation}\label{eq:multiangleeigenvalue}
\mu^{-1}= I_1 \pm \sqrt{I_0 I_2} \,.
\end{equation}
This is the multi-angle counterpart of our single-angle eigenvalue
equation of Eq.~(\ref{eq:singleangleeigenvalue}).

We introduce once more $\Omega=\gamma+\I\kappa$, and to split this
equation into its real and imaginary parts we write the integral
expressions in the form
\begin{equation}
I_n=J_n+\I K_n\,,
\end{equation}
where
\begin{eqnarray}
J_n&=&\int d\omega\,du\,g_{\omega,u}\,u^n\,
\frac{\omega+u\bar\lambda-\gamma}{(\omega+u\bar\lambda-\gamma)^2+\kappa^2}\,,
\nonumber\\
K_n&=&\int d\omega\,du\,g_{\omega,u}\,u^n\,
\frac{\kappa}{(\omega+u\bar\lambda-\gamma)^2+\kappa^2}\,.
\end{eqnarray}
Inserting this into Eq.~(\ref{eq:multiangleeigenvalue}) and equating
the real and imaginary parts separately yields the two real
equations
\begin{eqnarray}
(J_1-\mu^{-1})^2&=&K_1^2+J_0J_2-K_0K_2\,,
\nonumber\\
(J_1-\mu^{-1})&=&\frac{J_0K_2+K_0J_2}{2K_1}\,.
\end{eqnarray}
These two equations can now be taken as our consistency conditions,
analogous to Eqs.~(\ref{consistency-1}) and~(\ref{consistency-2}) in
the single-angle case.

Note that for a given angular mode $u=u_0$, the quantity $|Q_{\omega,u_0}|^2$
is a Lorentzian, centered at $\omega = \gamma - u_0 \bar\lambda$.
Thus, the range of $\omega$-modes that are affected by the instability
is different for different angular modes, as opposed to the 
single-angle case.

In order to solve for $\gamma$ and $\kappa$ satisfying the
consistency conditions, one may consider $\mu$ and $\bar\lambda$ as
independent quantities, keeping in mind the constraint $\lambda =
\bar\lambda - \epsilon \mu$. This allows us to find one real
equation from which $\mu$ is eliminated. As in the single-angle
case, this equation provides a $\mu$-independent relation between
$\gamma$ and $\kappa$ and thus the set of all possible eigenvalues.
The set of all points in the $\Omega=\gamma+\I\kappa$ plane
fulfilling this requirement represents the ``root locus diagram'' of
our self-consistency relations. The other equation is of the form
$\mu^{-1}=({\rm integral~expressions})$ and allows one to calculate
for any allowed $\Omega=\gamma+\I\kappa$ the corresponding $\mu$ by
quadratures alone.

The energy and angular distributions of SN neutrinos are
not independent of each other. On the other hand, in schematic
models one may assume that the angular distribution is independent
of energy and independent of flavor. In particular, the
approximation of a common blackbody neutrino sphere for all flavors
implies $g_{\omega,u}=g_\omega$, independently of $u$. In this case
the $u$ integration in the expressions for $I_n$ can be performed
explicitly, considerably simplifying the numerical inversion of the
self-consistency equations.

\subsection{Matter vs.\ neutrino background}
\label{matter-vs-neutrino}

Multi-angle effects on the stability of the spectrum may be
interpreted in terms of the separate effects of the two terms on the
r.h.s.\ of the EoM in Eq.~(\ref{eq:smallEoM}). The second term
represents the effects of Pantaleone's off-diagonal refractive index
caused by neutrino-neutrino interactions. It allows for the
instability in the first place and in addition can lead to
self-induced multi-angle
decoherence~\cite{Raffelt:2007yz,EstebanPretel:2007ec}. However,
this decoherence effect arises after some pendular oscillations and
would not be visible in the small-amplitude expansion. Decoherence
arises in a periodic system performing many revolutions and
different modes dephasing relative to each other. The unstable
solution in the small-amplitude expansion is not periodic, but
exponentially growing, and thus we can not study decoherence
effects. However, we can study the multi-angle modification
of the instability which is caused by $\bar\lambda$ in the first
term on the r.h.s.\ of Eq.~(\ref{eq:smallEoM}).

If Pantaleone's flavor off-diagonal refractive effect would not
exist, i.e.\ the last term on the r.h.s.\ in
Eq.~(\ref{stability-eom}) would be missing, then the first term on
the r.h.s.\ specifies that every mode $S_{\omega,u}$ freely
precesses around the weak-interaction direction with frequency
\hbox{$\omega_{\rm eff}=\omega+(\lambda+\epsilon\mu)u$}. We
recall that
\begin{equation}
\lambda=\sqrt{2}G_{\rm F}(n_e-n_{\bar e})\,\frac{R^2}{2r^2}\,,
\end{equation}
as given in Eq.~(\ref{lambdar-def}), represents the net matter
effect in the co-rotating frame. Note that we have kept only the
matter term causing an angle-dependent spread of $\omega_{\rm eff}$,
while the main matter effect, causing a common precession with
frequency $\sqrt{2}G_{\rm F}(n_e-n_{\bar e})$, has been removed by
going to a rotating frame.

The flavor-diagonal refractive effect caused by the
neutrino background, in our equations represented by $\epsilon\mu$,
plays a perfectly analogous role to ordinary matter. Indeed, it is
the quantity $\bar\lambda = \lambda + \epsilon \mu$ which appears
in the EoMs, and hence in the multi-angle analysis. Though we define
the neutrino background contribution in terms of the radial flux
densities and not their number densities, at large distances these
two quantities are identical. Therefore, we may write
\begin{equation}
\epsilon\mu=\sqrt{2}G_{\rm F}[(n_{\nu_e}-n_{\bar\nu_e})-
(n_{\nu_x}-n_{\bar\nu_x})]\,\frac{R^2}{2r^2}\, .
\end{equation}
The matter background term $\lambda$ and the neutrino background term
$\epsilon \mu$ are completely symmetric, if one takes into account
that the $x$ contribution is missing among charged leptons simply
because of the absence of charged $x$ leptons since both $\mu$ and
$\tau$ leptons are too heavy to be present in the SN context.

\subsection{Multi-angle suppression of the instability}
\label{sec:multi-suppression}

A large matter effect in the form of a large value of
$\bar\lambda=\lambda+\epsilon\mu$ can suppress the self-induced
instability in numerical SN simulations. This effect was predicted
in Ref.~\cite{EstebanPretel:2008ni} and has been repeatedly tested,
most recently in the numerical studies of
Ref.~\cite{Chakraborty:2011gd}. The reason invoked for this
suppression was the large dispersion of $\omega_{\rm
eff}=\omega+\bar\lambda u$ that is caused if $u$ is spread over the
unit interval and $\bar\lambda$ is much larger than the direct
spectral spread of $\omega$. It was then argued that this
multi-angle matter suppression of the instability would be
noticeable if the matter term were much larger than the neutrino
background term, i.e.\ if $\lambda\gg \mu$.  The linearized analysis
allows us to show analytically that when $\bar\lambda\gg\mu$, the
collective oscillations are completely suppressed.

Let $\omega_{\rm max}$ be the largest value of $|\omega|$ where
$g_{\omega,u}$ is significant. When $\bar\lambda \gg \omega_{max}$,
the quantities $I_n$ in Eq.~(\ref{In-def}) are suppressed due to the
factor of $\bar\lambda u$ in the denominator. The cancellation of
large values of $\bar\lambda u$ by $\Omega$ is possible only in a
narrow range of $u$ values, which will give only a small
contribution to the integral.
Equation~(\ref{eq:multiangleeigenvalue}) can then be satisfied only
at $\mu \sim \bar\lambda$. Therefore, the system should be
stable for $\bar\lambda \gg \mu$, i.e.~when the net effects of
ordinary matter dominate over those of the neutrino-neutrino
interactions.

A seemingly different multi-angle suppression affects the
``saturation mode'' of collective oscillations that can occur in
spectra with at least two zero crossings for very large $\mu$
(Secs.~\ref{three-box} and~\ref{four-box}). This mode was first
discovered in an analytic toy model~\cite{Raffelt:2008hr} and later
re-discovered in the context of a more realistic SN
example~\cite{Duan:2010bf}. Moreover, both papers argued and tested
numerically that the spread of $\omega_{\rm eff}$ caused by the
neutrino-neutrino term would suppress the saturation mode. This
suppression effect, however, is just another aspect of the impact of
$\bar\lambda$ in the EoMs. The matter and neutrino backgrounds
together cause a common multi-angle effect by
dispersing~$\omega_{\rm eff}$. This may be seen as follows.

In the single-angle approach and for spectra with $\epsilon$ not
very small, the consistency relations imply that $\kappa$ cannot be
much larger than a typical spread $\Delta\omega$ of the frequency
spectrum. Indeed, for $\kappa \gg \Delta\omega$,
Eq.~(\ref{consistency-1}) would not be satisfied for any finite
asymmetry $\epsilon$. Assuming that $\kappa$ and thus the resonance
width remains of order $\Delta\omega$ in the multi-angle case, it is
clear that a large dispersion of $\omega_{\rm eff}$ shifts most
modes away from the resonance. In this way, multi-angle effects
would indeed suppress self-induced flavor conversions at large
$\bar\lambda$. We will see that the presence of a large
$\bar\lambda$ in the EoMs will never increase $\kappa$
significantly, although it can suppress it considerably. In this
sense the reasons given here for the multi-angle matter suppression
indeed apply.

For certain spectra, notably the generic two-box case, the
single-angle approximation predicts a cutoff $\mu_{\rm sync}$. The
instability cannot form at large values of $\mu$ since the
consistency conditions in Eqs.~(\ref{consistency-1}) and
(\ref{consistency-2}) cannot be satisfied for any complex $\Omega$,
and the system is stable. For $\mu < \mu_{sync}$, however, pendular
oscillations commence. The multi-angle analysis does not change this
behavior qualitatively. The value of $\mu_{\rm sync}$ in the
multi-angle case is some average of the $\mu_{\rm sync}$ values for
different $u_0$ values, as can be seen in
Fig.~\ref{fig:2box-single}.

\subsection{Normal vs.\ inverted mass hierarchy}
\label{hierarchy-multi}

In the single-angle approximation, we saw that as far as the
stability analysis is concerned, analysis of normal hierarchy is the
same as that of the inverted one, except for a change of the sign of
$\mu$. This is true also in the multi-angle scenario, except that
one also needs to change the sign of $\lambda$. Indeed, normal
hierarchy changes Eq.~(\ref{eq:smallEoM}) to
\begin{eqnarray}
\label{eq:smallEoM-normal}
\I\partial_r \tilde{S}_{\omega,u}
&=&\left[- \omega+u(\lambda+\epsilon\mu)\right]\tilde{S}_{\omega,u}
\nonumber\\
&-&\mu \int du'\,d\omega'\,(u+u')\,g_{\omega'u'}\, \tilde{S}_{\omega',u'}\,.
\label{stability-eom-normal}
\end{eqnarray}
The solution of this equation can be given in terms of
the solution $S_{\omega,u}$ of Eq.~(\ref{eq:smallEoM}) as
\begin{eqnarray}
\tilde{S}_{\omega,u}(\mu,\lambda,g_{\omega,u})
& = & S^*_{\omega,u}(\mu,-\lambda, -g_{\omega,u})  \nonumber \\
& = &  S^*_{\omega,u}(-\mu, -\lambda, g_{\omega,u}) \; .
\end{eqnarray}
Since $S$ and $S^*$ should have the same stability behavior,
this implies that the stability conditions for normal hierarchy
are the same as those for the inverted hierarchy with a change in the
sign of $g_\omega$ or $\mu$ (not both at the same time), and
an additional change in the sign of $\lambda$.

\section{Multi-angle examples}
\label{sec:multi-ex}

We now study the stability conditions including multi-angle effects
for the box spectra considered earlier as well as for a realistic SN
example. It turns out that the multi-angle effects modify the
single-angle results in significant ways. We always take the
emission to be uniform over $0 \leq u \leq 1$. The integrals $J_n$
and $K_n$ can then be analytically calculated for the box spectra.
The expressions, however, are not very illuminating, and we do not
give them here.

\subsection{Two boxes}
\label{two-box-multi}

The two-box spectrum and the single-angle eigenvalues $\gamma(\mu)$
and $\kappa(\mu)$ for different choices of the emission angle $u_0$
were shown in Fig.~\ref{fig:2box-single}. In addition, we show in
Fig.~\ref{fig:2box-single} the numerical result for $\gamma(\mu)$
and $\kappa(\mu)$ in the multi-angle case, assuming the absence of
matter so that $\bar\lambda=\epsilon\mu$ and all effects are caused
by neutrino-neutrino interactions alone. As far as the stability
behavior is concerned,  the multi-angle effects may be interpreted
as some (complicated) average of the single-angle effects with
different $u_0$ values. In particular, the value of $\kappa$ at any
value of $\mu$ lies within the range of $\kappa$ for different $u_0$
values in the single-angle approximation. In other words,
multi-angle effects leave the system qualitatively unchanged.

\begin{figure}
\includegraphics[width=0.8\columnwidth]{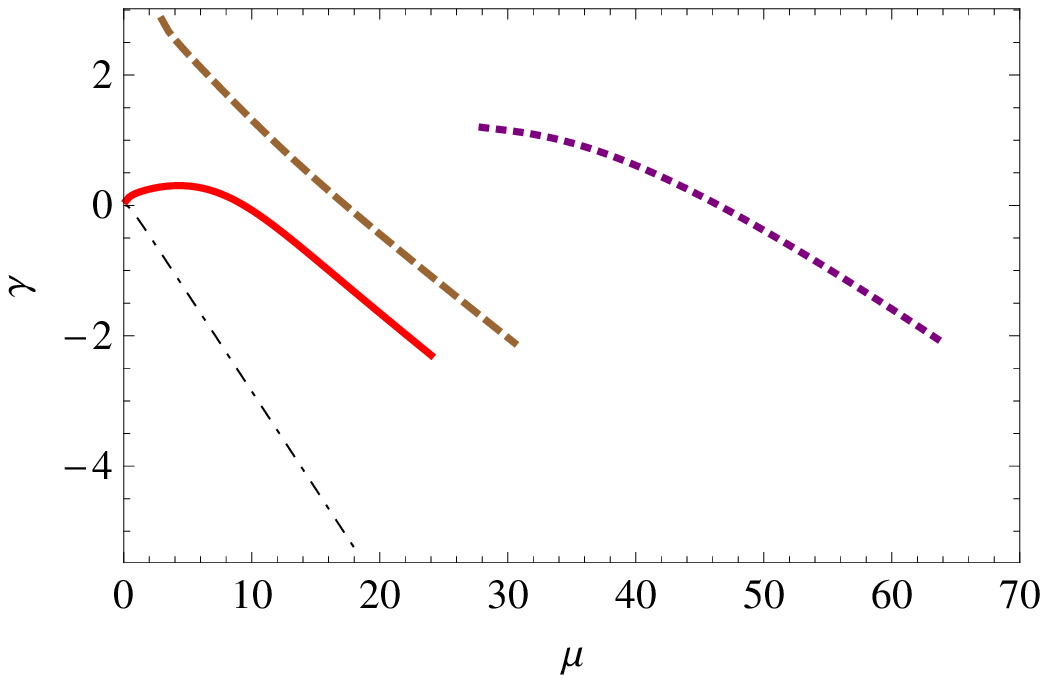}
\vspace{0.1in}
\includegraphics[width=0.8\columnwidth]{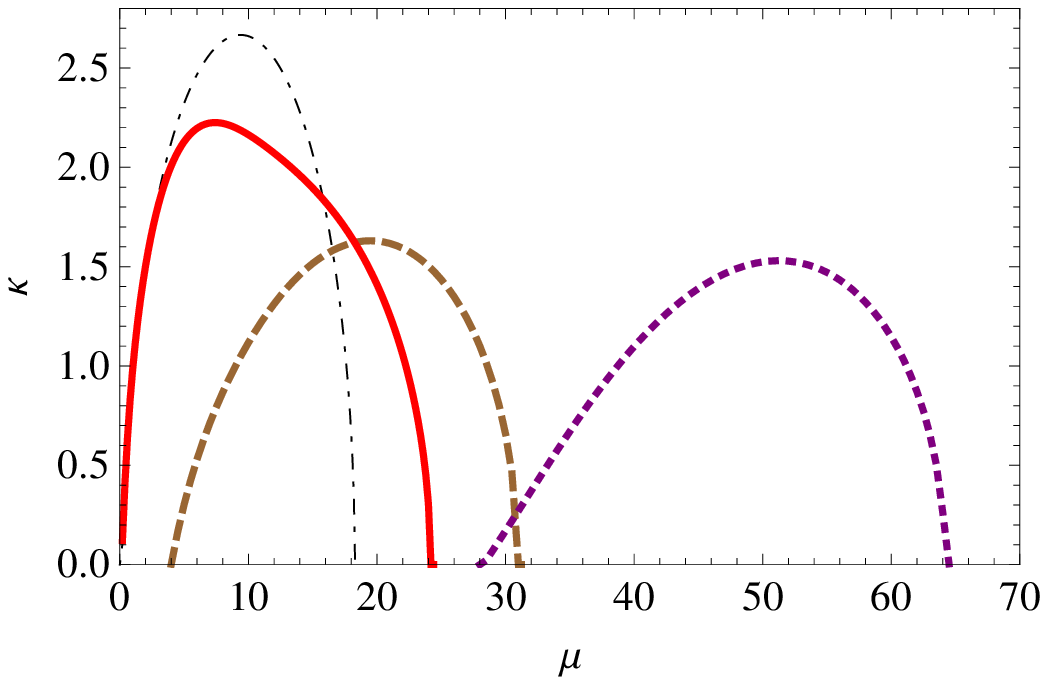}
\caption{Multi-angle eigenvalues $\gamma(\mu)$ and $\kappa(\mu)$ for
the two-box spectrum in
Fig.~\ref{fig:2box-single}.
Black (dash-dotted) thin line: single-angle ($u_0=1/2$). All other
lines: multi-angle. Red (solid): no matter ($\lambda=0$). Brown
(dashed): $\lambda=5$. Purple (dotted): $\lambda=30$. The
$\gamma(\mu)$ curves are only shown where $\kappa(\mu)\not=0$.
\label{fig:2box-multi}}
\end{figure}

For sufficiently dense matter the picture changes considerably. In
Fig.~\ref{fig:2box-multi} we repeat the single-angle results for
$u_0=1/2$ (thin black dot-dashed curves) and the no-matter
multi-angle results (red solid curve). We also show curves for
$\lambda=5$ and 30. Qualitatively the $\gamma(\mu)$ and
$\kappa(\mu)$ curves are shifted to larger values of $\mu$, the
spectra being stable below some threshold value and above some
cutoff value. The interval of unstable $\mu$ values remains roughly
the same, except that it is shifted to larger $\mu$ values. The
shift is approximately such that the unstable domain arises for
$\mu\sim\bar\lambda$ as predicted earlier.

The presence of matter alone does not stabilize the system, it
shifts the instability domain to larger $\mu$ values. Indeed, a
nonzero $\lambda$ can allow for an instability even at the values of
$\mu$ where $\epsilon \mu$ alone could not have generated an
instability. In the $\mu$-$\lambda$ plane, the system is unstable
along a strip roughly following $\lambda\sim\mu$, and stable
outside. In a SN we sweep from large to small $\mu$ values as a
function of radius, and at the same time from large to small
$\lambda$ values. If everywhere along the path $\lambda\gg\mu$, we
would be outside the instability strip.

However, even if this is not the case, the system could still be
stabilized. We also have the effect of dispersing the spectrum
relative to the resonance condition. In a spectrum of given spectral
width $\Delta\omega$, the resonance denominator
$\omega+u\bar\lambda-\Omega$ will be on resonance only for a narrow
range of $u$ values if $\bar\lambda$ is large. Therefore, even if
the system is not stable, most modes will be off-resonance in
analogy to our discussion of spectral tails.

A detailed understanding of the multi-angle stabilization of the
neutrino flux streaming from a SN core as studied most recently in
Ref.~\cite{Chakraborty:2011gd} requires more detailed scrutiny
because the described effects of shift and dispersion both affect a
realistic SN neutrino flux.

\subsection{Three boxes}
\label{three-box-multi}

\begin{figure}
\includegraphics[width=0.8\columnwidth]{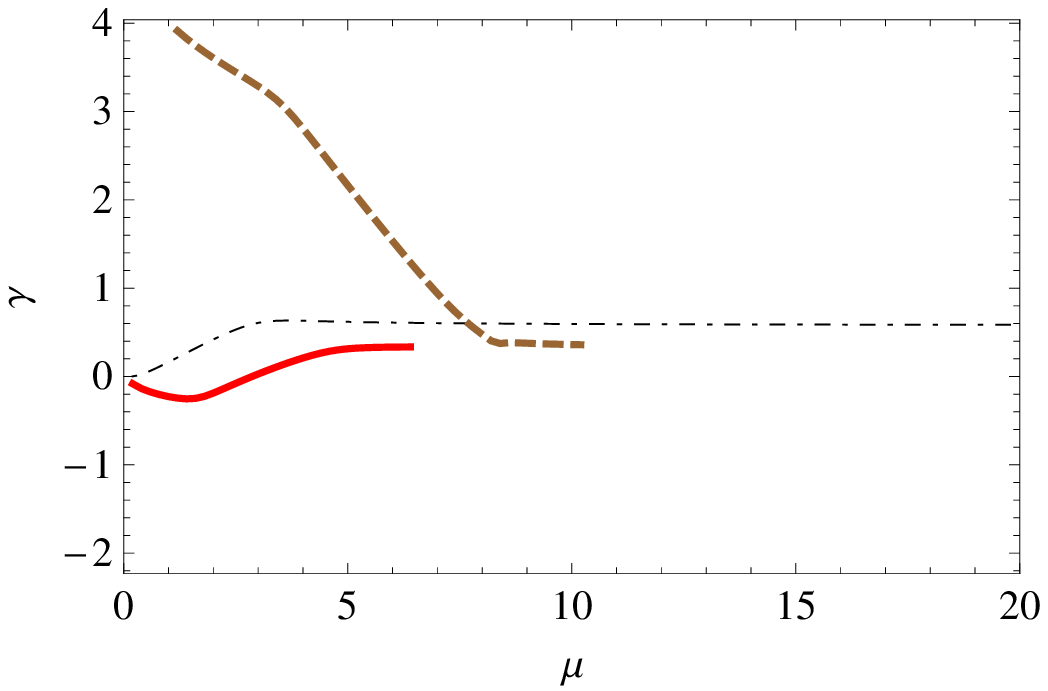}
\includegraphics[width=0.8\columnwidth]{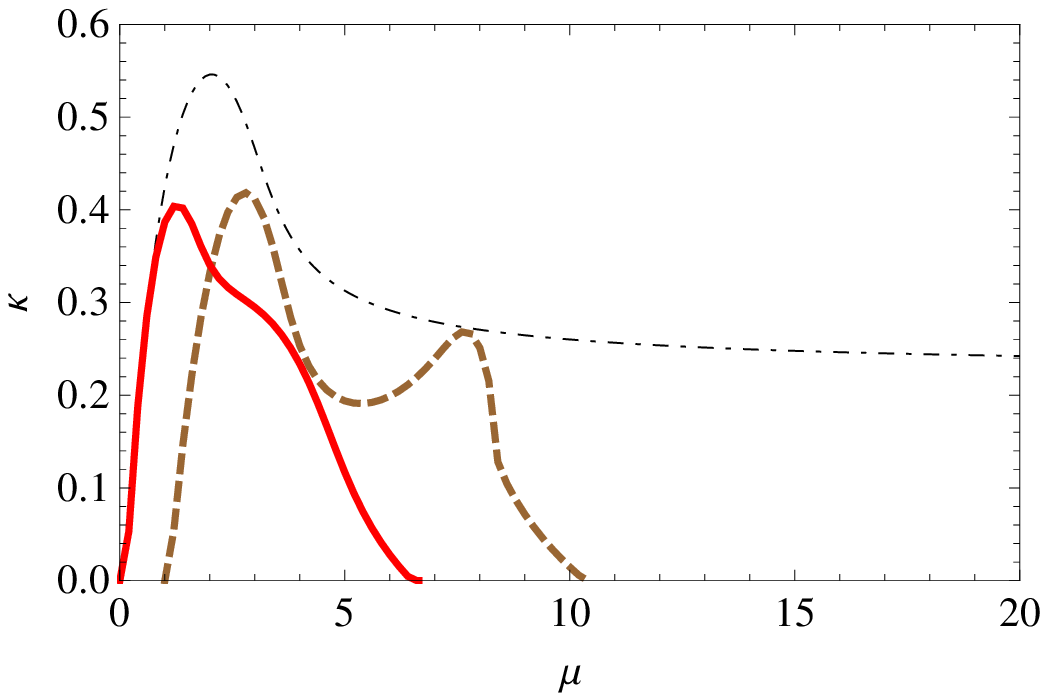}
\caption{Multi-angle eigenvalues $\gamma(\mu)$ and $\kappa(\mu)$ for
the three-box spectrum in
Fig.~\ref{fig:3box-single}.
Black (dash-dotted) thin line: single-angle ($u_0=1/2$). All other
lines: multi-angle. Red (solid): no matter ($\lambda=0$). Brown
(dashed): $\lambda=5$. The
$\gamma(\mu)$ curves are only shown where $\kappa(\mu)\not=0$.
\label{fig:3box-multi}}
\end{figure}

We now turn to the three-box example of Fig.~\ref{fig:3box-single}
which was chosen such that we observe a ``saturation mode:'' The
system is unstable with an asymptotic value of $\kappa$ for
arbitrarily large $\mu$ and the center of the resonance $\gamma$
remains centrally located in the spectral range. Here the
multi-angle matter effect causes a dramatic modification of the
single-angle results because the curve $\kappa(\mu)$ is completely
suppressed above a critical $\mu$-value, i.e.\ the spectrum is
actually stabilized.

In Fig.~\ref{fig:3box-multi} we repeat the single-angle case with
$u_0=1/2$ and show the multi-angle case with vanishing matter so
that $\bar\lambda=\epsilon\mu$. The previous saturation effect is
completely suppressed. As we increase $\lambda$, the instability
domain is shifted to larger $\mu$ values roughly linearly with
$\lambda$ as explained earlier. In other words, the system now
behaves similar to the two box case.

\subsection{Four boxes}
\label{four-box-multi}

The main novelty of the four-box spectrum is that it can display two
simultaneous instabilities. One of them can be of the saturation
type, similar to the three-box example, details depending on the
exact choice of parameters. Multi-angle effects for $\lambda=0$ and
thus $\bar\lambda=\epsilon\mu$ are shown in
Fig.~\ref{fig:4box-single}. The results are entirely in line with
the expectations from the previous examples in that the saturation
mode gets suppressed at large $\mu$. We shall not repeat the
analysis with the four-box spectra here. However the realistic SN
spectrum in the next section can be seen to be close to a four-box
spectrum, and is observed to display the same expected behavior.

\subsection{A realistic SN spectrum}
\label{sn-multi}

We now apply our stability analysis to a realistic SN spectrum that
is motivated by Duan and Friedland's recent numerical study of
multi-angle suppression~\cite{Duan:2010bf}. They used $\nu_e$,
$\bar\nu_e$, and $\nu_x$ spectra with assumed Fermi-Dirac form. The
temperatures were taken to be 2.1, 3.5, and 4.4~MeV, respectively,
and the degeneracy parameters 3.9, 2.3 and 2.1. The average energies
are then 9.4, 13.0 and 15.8~MeV. The ratios of number fluxes are
taken to be $1.3 : 1.0 : 1.5$, where the total $\bar\nu_e$ flux is
normalized to unity as per our convention introduced in
Sec.~\ref{hamiltonian}.  We use $\Delta m^2 = (50 {\rm meV})^2$ in
order to convert the energy scale to $\omega$, which we show in the
units of km$^{-1}$.

\begin{figure}[h!]
\includegraphics[width=0.8\columnwidth]{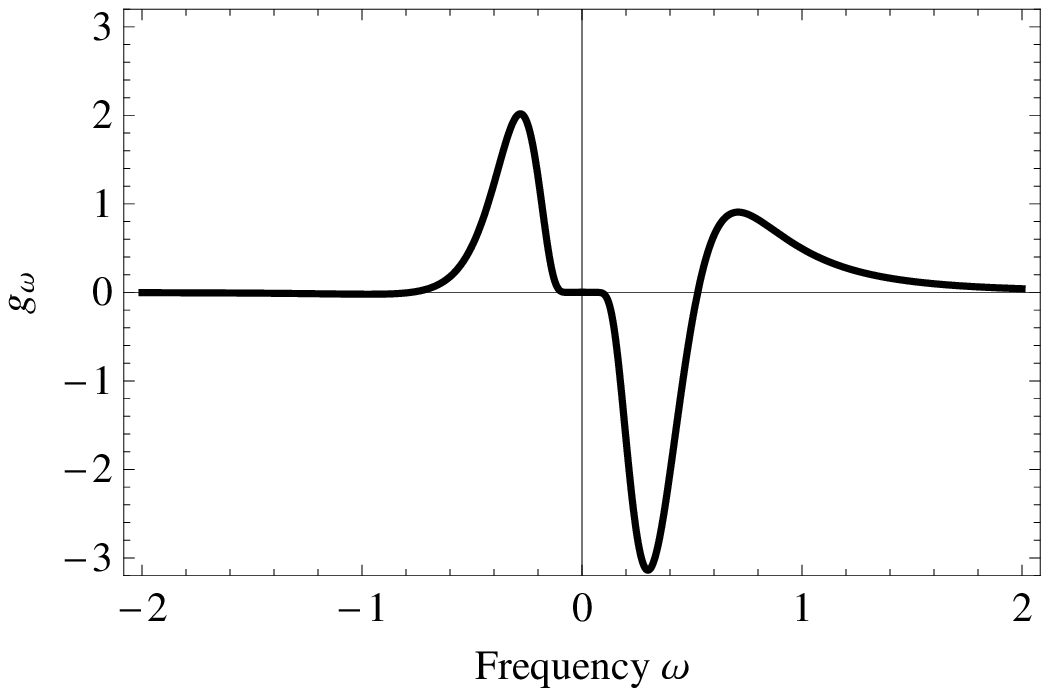}
\vspace{0.2in}
\includegraphics[width=0.8\columnwidth]{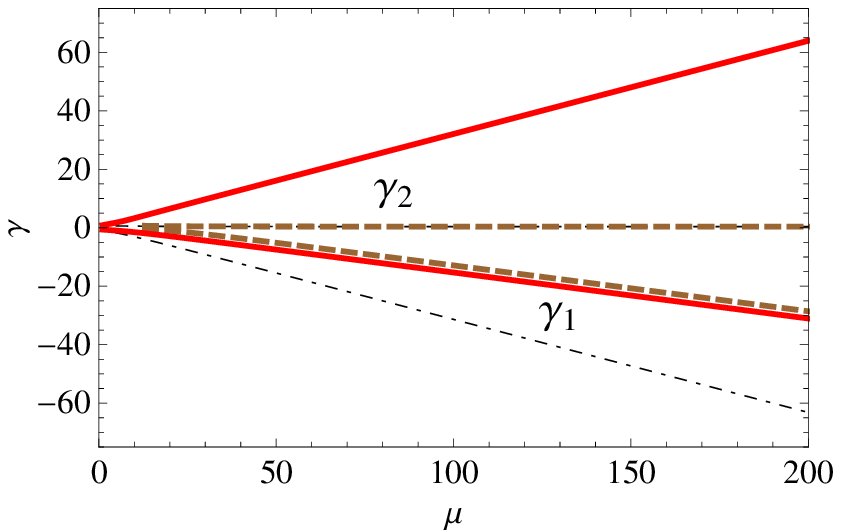}
\vspace{0.1in}
\includegraphics[width=0.8\columnwidth]{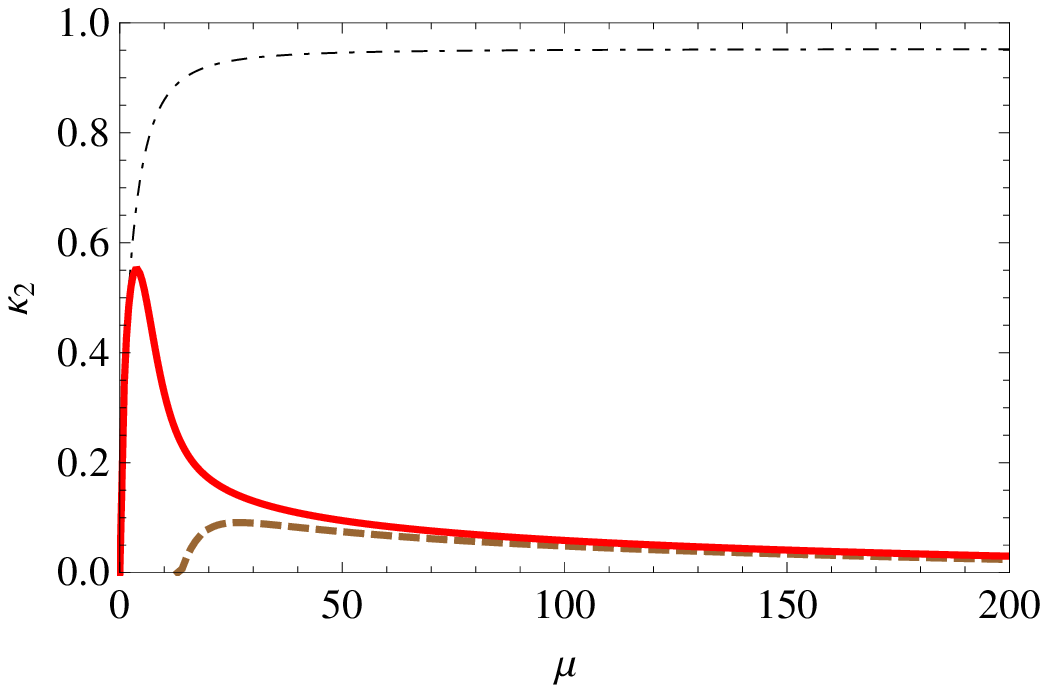}
\vspace{0.1in}
\includegraphics[width=0.8\columnwidth]{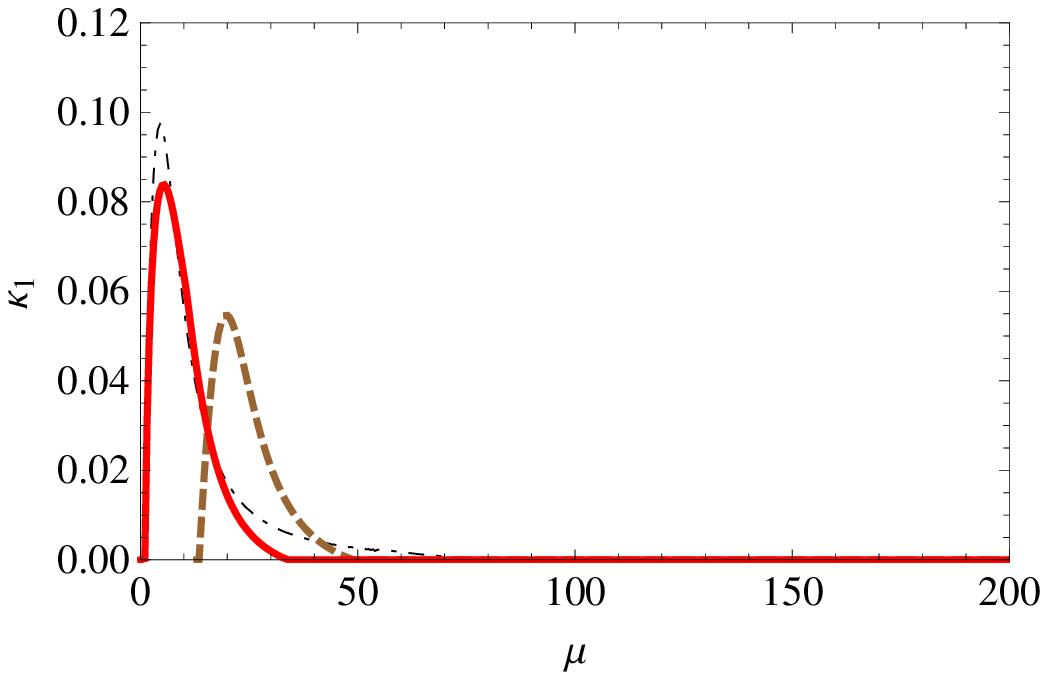}
\caption{Realistic SN spectrum as described in the text and identical
with Ref.~\cite{Duan:2010bf}. Lower panels: $\gamma(\mu)$ and $\kappa(\mu)$
for the two instabilities, where 1 denotes the one with smaller~$\gamma$.
Mode~2 is the saturation mode.
Black (dash-dotted) thin line: single-angle ($u_0=1/2$),
otherwise multi angle. Red (solid): no matter ($\lambda=0$).
Brown (dashed): $\lambda=5$.
\label{fig:sn-multi}}
\end{figure}

The spectrum is shown in the top panel of Fig.~\ref{fig:sn-multi}.
It has three spectral crossings and thus compares with a four-box
spectrum, although the deviation from zero of the left-most part is
almost invisible.  (For $\omega < -0.8$, the value of $g_\omega$ is
small and negative.) The main feature of interest here and in Duan
and Friedland's study is the presence of a ``saturation mode'' in
which the instability in the single-angle treatment exists for
arbitrarily large $\mu$. Such a feature requires at least two
crossings, so in this regard the visual impression of this being a
``three box'' example is actually the main point.

In single angle (thin dot-dashed black line) we find a dominant
``saturation mode'' as expected (corresponding to $\gamma_2,
\kappa_2$) and in addition a ``cutoff mode'' (corresponding to
$\gamma_1, \kappa_1$) which always has very small values for
$\kappa(\mu)$. The existence of this mode is a residue of the not
perfect disappearance of $g_\omega$ at $\omega\alt-0.8$, i.e.\ of
the small ``fourth box.'' If we were to truncate the spectrum at the
left-most crossing, this mode would disappear. At very small $\mu$,
these two instabilities are centered near the positive spectral
crossings as expected. Even the ``cutoff mode'' does not completely
disappear for large $\mu$ due to the spectral tails.

Next we include multi-angle effects without matter ($\lambda=0$),
roughly corresponding to the study of Ref.~\cite{Duan:2010bf} where
the matter density was small compared to the neutrino density.
The red solid curve represents the case with
$\bar\lambda=\epsilon\mu$. We see that at large $\mu$, corresponding
to small radii in the SN, the saturation mode is indeed suppressed.
It does not vanish completely, owing to the presence of spectral
tails, in contrast to the earlier box examples. The center of the
saturation mode, $\gamma_2 - u \epsilon \mu$, continues to be in the
central part of the spectrum for $u \approx 1$. The center of the
cut-off mode, $\gamma_1 - u \epsilon \mu$, on the other hand, is
actually pushed into the tail of the spectrum.

Next we add a significant amount of matter, represented by
$\lambda=5$ (brown dashed curves). There now a lower $\mu$-threshold
appears for both, the cutoff mode as well as the saturation mode.
This is a consequence of the complete suppression of collective
oscillations at $\bar\lambda \gg \mu$, predicted in
Sec.~\ref{sec:multi-suppression}. Thus, our analytical prediction is
vindicated in this scenario of a realistic SN spectrum.

The main result of Ref.~\cite{Duan:2010bf}, obtained in the regime
$\bar\lambda \sim \mu$, was that multi-angle effects stabilize the
spectrum in deep layers of the SN where in single-angle strong
bipolar oscillations would have occurred. Our stability analysis
provides two reasons for this behavior. The value of $\kappa(\mu)$
is strongly reduced, although the spectrum is always unstable as
behooves a spectrum with tails. Moreover, the dispersion of
effective oscillation frequencies due to the presence of
$u\bar\lambda$ in the resonance denominator moves most modes off
resonance. Given $\mu(r)$ and $\lambda(r)$, one can predict the
exact SN radius where serious bipolar oscillations would kick in for
the multi-angle case, with the stability analysis.

\section{Multi-angle instability in normal hierarchy}
\label{sec:novel}

A perfectly symmetric spectrum shows self-induced
kinematical decoherence among angular modes in either hierarchy, or
in the language of our present study, whether the zero-crossing of
$g_\omega$ is positive or negative~\cite{Raffelt:2007yz}. The same
is found for spectra with a small asymmetry $\epsilon\ll 1$
\cite{EstebanPretel:2007ec}. Kinematical decoherence requires the
spectrum to be unstable in the first place, so these findings imply
that in the multi-angle case, an instability does not require a
positive zero-crossing of $g_\omega$. This surprising result is
easily verified with our method.

Let us assume that $\bar\lambda=0$, as would be the case for an
antisymmetric spectrum ($\epsilon=0$) and in the absence of matter.
We further make the simplifying assumption of a universal angular
distribution, i.e.\ $g_{\omega,u} = g_\omega f_u$, with $\int du
f_u =1$. Let us introduce the notation
\begin{equation}
G=\int d\omega\,\frac{g_\omega}{\omega-\Omega}\,.
\end{equation}
Then in the multi-angle case, we have
\begin{equation}
I_n=G\int_0^1 du\, u^n = G \langle u^n \rangle \; .
\end{equation}
The eigenvalue equation is then
\begin{equation}
\mu^{-1}=I_1\pm\sqrt{I_0 I_2}=
\left[\langle u\rangle\pm\langle u^2\rangle^{1/2}\right]\,G\,.
\end{equation}
The stability analysis then corresponds to determining the
solutions $\Omega = \gamma \pm i \kappa$ for the equation
\begin{equation}
\mu^{-1} = {\cal K}_\pm G \; ,
\end{equation}
where ${\cal K}_\pm \equiv \langle u\rangle\pm\langle u^2\rangle^{1/2}$.

In the single-angle approximation, $\langle u \rangle = u_0,
\langle u^2 \rangle = u_0^2$, so that ${\cal K}_+ = 2u_0$ and
${\cal K}_- = 0$. The latter equation has no solution, since it
would require $\mu^{-1}=0$ identically. One may then write
\begin{equation}
\mu^{-1}_{\rm single} = 2 u_0 G \; .
\end{equation}
This may be solved to obtain the values of $\gamma$ and $\kappa$ in
the single-angle approximation. Clearly, $\mu^{-1}_{\rm single}$ has
the same sign as $G$, so that $G$ is positive for a spectrum with
positive crossing.

We can now write the two values of $\mu^{-1}$ in multi-angle
analysis that correspond to the same values of $\gamma$ and
$\kappa$:
\begin{equation}
\mu^{-1}_+  =  {\cal K}_+ G \; , \quad
\mu^{-1}_- =  {\cal K}_- G \; ,
\end{equation}
Thus, there are two values of $\mu$ in multi-angle analysis that
correspond to the same $(\gamma,\kappa)$ in the single-angle
analysis. The multi-angle stability behavior at these two $\mu$
values can then be understood in terms of the single-angle stability
behavior at $\mu_{\rm single}$.

Note that ${\cal K}_+ \geq 0$ and ${\cal K}_- \leq0$. Therefore,
$\mu_+$ is positive, and it corresponds to just a scaled value of
$\mu_{\rm single}$ with the same (inverted) hierarchy. The
single-angle study is indeed a good proxy for the multi-angle case
in such a scenario. On the other hand, $\mu_-$ is negative, and
corresponds to the normal hierarchy. For spectra with $\langle
u^2\rangle^{1/2}\sim\langle u\rangle$ one finds that $|\mu_-|\gg
|\mu_+|$, so the ``wrong hierarchy'' case shows the same instability
at much larger densities.

The exact mapping between the single- and multi-angle cases is not
possible for the more realistic case of $\bar\lambda$ being small
but nonzero, but of course we expect a qualitatively similar
behavior.

In Ref.~\cite{EstebanPretel:2007ec} it was found that kinematical
decoherence is suppressed for sufficiently large $\epsilon$, which
in normal hierarchy means that the spectrum is stable and the
``positive crossing'' rule for an instability is back in force.
These studies did not include the presence of matter, so
$\bar\lambda=\epsilon\mu$ with a sufficiently large $\epsilon$
implied that $\bar\lambda$ was not very small. In this sense, the
stability in normal hierarchy for sufficiently large $\epsilon$
looks like yet another case of multi-angle suppression of an
instability.

\section{Conclusions and outlook}
\label{sec:conclusions}

The refractive effect of neutrino-neutrino interactions can cause a
flavor instability among the different energy and angular modes,
exchanging the flavor content while leaving unchanged the overall
flavor content of the entire ensemble. In the context of SN
neutrinos, the focus so far has been on impacts of collective
flavor oscillations on the neutrino signal of the next nearby SN or
for the r-process nucleosynthesis. Many analyses have been performed
under the simplifying assumption that all neutrinos feel
the same refractive effect due to the other neutrinos (single-angle
assumption). However it has also been found that multi-angle
effects can strongly modify the answer and can typically suppress
flavor conversions that would occur in the single-angle treatment.
In this sense, a more pressing question is to understand the
stability conditions for a dense neutrino stream, and to understand
if the usual neglect of flavor oscillations in those dense SN
regions is justified where a Boltzmann treatment of neutrino
transport is necessary.

We have studied the conditions under which instabilities can occur
in the neutrino stream in the two-flavor framework. While the
equations of motion are nonlinear, the run-away solutions can be
found by a standard mode analysis of the linearized system, i.e.\ in
the small-amplitude expansion. We have shown how the stability
question is reduced to a straightforward eigenvalue equation in the
co-rotating frame. The existence of complex eigenvalues
directly corresponds to an instability. In the multi-angle case,
besides the small-amplitude expansion, we have linearized the
equations in the small angular divergence of the neutrino flux that
is applicable at a large distance from the source. In the
single-angle case, our conditions agree with the previous
literature, whereas in the multi-angle case, our results are new.

In the single-angle case and using schematic energy spectra (box
spectra in the $\omega$ variable), the eigenvalue equation can be
solved analytically. We identify two generic instability cases. The
``cutoff mode'' exists for a finite range of neutrino densities, or
equivalently, a finite interval of effective neutrino-neutrino
interaction strengths $\mu_0<\mu<\mu_1$. In particular, the system
is stable above a cutoff value $\mu_1$, also termed the
``synchronization'' or ``sleeping top'' regime. This behavior is
generic to a single-crossed (or two-box) spectrum and represents the
traditional flavor pendulum. On the other hand, there is the
``saturation mode'' where the instability approaches an asymptotic
value for large $\mu$. This mode requires at least two crossings
(three boxes) and had been previously found in the literature.
When it exists, the saturation mode exists for both
hierarchies. For multi-crossed spectra, several modes can co-exist,
for example in the four-box case a saturation mode and a cutoff mode
or two cutoff modes.

An additional case occurs for neutrino spectra with a vanishing
asymmetry, i.e.\ with no net lepton number flux. The run-away rate
$\kappa$ grows indefinitely as a function of $\mu$. This actually is
a special case of the cutoff mode (cutoff at infinite $\mu$) and
does not seem to be of practical interest in the SN context.

Realistic neutrino spectra are not described by boxes and have tails
as a function of $\omega=|\Delta m^2|/2E$, corresponding to infrared
energies. Such spectra do not have a cutoff and a synchronization
regime does not exist. However, for large $\mu$ the instability
moves entirely into the infrared regime and has no practical impact.
There remains an effective cutoff density where the amount of flavor
exchange between modes drops by orders of magnitude within a narrow
range of neutrino densities. Still, the existence of a
synchronization regime is only an approximate concept.

Multi-angle effects modify the instability in several ways. For
example, while the single-angle approximation predicts an
instability only if there is a positive zero-crossing of the
spectrum $g_\omega$, the multi-angle analysis implies that spectra
with a small asymmetry (a small net lepton number flux) and in the
absence of matter are unstable in both hierarchies, whatever be the
sign of the zero-crossing. The solutions for both hierarchies can be
mapped onto each other.

The main multi-angle modification of the instability is caused by a
background of matter, where here neutrinos themselves also
contribute to the matter effect. The saturation mode is ``fragile''
and, for box spectra, develops a cutoff. (For realistic spectra with
tails, there is no exact cutoff, however, as mentioned earlier.) The
neutrino background alone is enough to achieve this effect.

For both the saturation and cutoff modes, a large matter effect
shifts the domain of instability $\mu_0<\mu<\mu_1$ to larger values
of $\mu$. The amount of shift is approximately equal to the matter
potential $\lambda$, apparently without reducing the
typical $\kappa$ values within the instability domain. An
additional consequence is that the affected range of modes shrinks
for increasing $\lambda$. Even if the run-away scale
$\kappa$ is not reduced, only a narrow range of angular modes is on
resonance if $\lambda$ is large. For realistic spectra with tails,
$\kappa(\mu)$ does not vanish for large $\mu$, i.e.\ there is not an
exact cutoff. On the other hand, there is a lower threshold
$\mu_0\sim\lambda$ below which the spectrum is completely stable. In
other words, multi-angle effects introduce a range
$0<\mu<\mu_0\sim\lambda$ where the spectrum is stable even when such
a stable range does not exist in the single-angle approximation.
This behavior corresponds to the multi-angle suppression of
instability at $\lambda \gg \mu$, which we motivate analytically
through our linearized treatment.

One may further investigate these findings in the context of
realistic SN situations where neutrinos stream from regions of large
density to vacuum, sweeping through $\mu$ values from very large to
zero. The central frequency $\gamma$ of the instability then sweeps
through different spectral ranges and the growth rate $\kappa$
depends on both the neutrino and matter density. Effective flavor
conversion can be suppressed in different ways: The growth rate
$\kappa$ can vanish when $\mu$ falls into the multi-angle-implied
stability domain, or it can be too small on the available distance
scale, or the instability can be located mostly in the infrared and
affect only an ignorable part of the spectrum.

We have considered neutrinos streaming from a source and have made
the approximation of small angular divergence that is appropriate at
a large distance from the source. Our approach can be easily adapted
to a homogeneous ensemble with a nontrivial angular distribution
that evolves in time rather than as a function of distance from a
source. In this case no expansion in the angular variable is needed.

We have only considered trivial angular distributions where all
neutrino flavors are assumed to be emitted from a common blackbody
sphere without limb darkening, corresponding to a box spectrum in
our $u$ variable. One can extend our analysis to more complicated
angular distributions that could involve, for example, spectral
crossings in the angular variable. As advocated by Sawyer, such
crossings cause a ``multi-angle instability,'' but its practical
relevance remains to be investigated.

We have made the usual assumption of azimuthal symmetry of the
neutrino flux around the radial direction away from the source. In
the long-distance limit of small angular divergence of the neutrino
field, one can of course expand the EoMs in terms of small variables
that depend on two angles. It remains to be explored if cylindrical
symmetry enforces unphysical solutions, just as the single-angle
approximation sometimes enforces unphysical solutions, and if there
are new instabilities or important modifications that could be
relevant in the SN context. Based on schematic models using a
small number of modes it has been suggested that violations of
cylindrical symmetry could have a major impact~\cite{Sawyer:2008zs}.

A rich phenomenology of neutrino flavor conversions is bound to
emerge from further investigations into the collective effects due
to neutrino-neutrino interactions and ordinary matter background, in
the complete multi-angle framework.

\section*{Acknowledgements} 

We would like to thank Basudeb Dasgupta, Srdjan Sarikas and Ray Sawyer 
for their comments on the manuscript.
In Munich, this work was partly supported by the Deutsche
Forschungsgemeinschaft under grants TR-27 and EXC-153. G.R.\ thanks
the Tata Institute of Fundamental Research (Mumbai) for hospitality
while this work was begun.


\end{document}